\documentclass[journal, twocolumn]{IEEEtran}

\ifCLASSINFOpdf
\else
\fi
\newcommand{\subparagraph}{}

\usepackage{amsmath}
\usepackage{amssymb}
\usepackage{mathtools}
\usepackage[noadjust]{cite}
\usepackage{soul}
\usepackage{enumerate}
\usepackage{graphicx}
\usepackage{blkarray}
\usepackage{float}
\usepackage[dvipsnames]{xcolor}
\usepackage{pgfplots}
\usepackage{caption}
\usepackage{subcaption}
\usetikzlibrary{intersections}
\pgfplotsset{compat=1.13} 
\usepackage{lipsum} % for generating filler text
\usepackage{titlesec}

\titleformat{\subsection}[runin]% runin puts it in the same paragraph
        {\normalfont\bfseries\itshape}% formatting commands to apply to the whole heading
        {\thesubsection}% the label and number
        {0.5em}% space between label/number and subsection title
        {}% formatting commands applied just to subsection title
        [.]% punctuation or other commands following subsection title

\DeclarePairedDelimiter{\ceil}{\lceil}{\rceil}
\DeclarePairedDelimiter{\floor}{\lfloor}{\rfloor}

\def\str#1{{\mathcal #1}}
\newcommand{\dfb}{\stackrel{\Delta}{=}}

\newcommand{\Z}{\mathbb{Z}}
\newcommand{\X}{\mathcal{X}}
\renewcommand{\S}{\mathcal{S}}
\renewcommand{\O}{\mathcal{O}}
\newcommand{\I}{\mathcal{I}}

\newtheorem{theorem}{Theorem}

\newtheorem{proposition}{Proposition}

\newtheorem{lemma}{Lemma}

\newtheorem{corollary}{Corollary}

\newtheorem{case}{Case}

\makeatletter
\@addtoreset{case}{subsection}
\makeatother

\pgfplotsset{
    standard/.style={
        axis x line=middle,  
        axis y line=middle,
        enlarge x limits=0.15,
        enlarge y limits=0.15,
		every axis x label/.style={at={(current axis.right of origin)},anchor=north west},
		label style={font=\small},
		tick label style={font=\small},
		every axis y label/.style={at={(current axis.above origin)},anchor=north east}
		label style={font=\small},
		tick label style={font=\small}		 
	}
}

\title{Evolutionary Matrix-Game Dynamics Under Imitation in Heterogeneous Populations}
\author{Yiheng Fu and Pouria Ramazi%
\thanks{Y. Fu and P. Ramazi are with
        Faculty of Science
        University of Alberta, 
	Canada,
          {\tt\small yiheng5@ualberta.ca,\,p.ramazi@gmail.com}.
    }
    \thanks{
This paper is published in part in the proceedings of the 2019 IEEE Conference on Decision and Control, Nice, France, December 11-13, 2019 \cite{fu2019asynchronous}. 
This paper additionally includes 
Section \ref{sec:example} on numerical examples, 
Proposition \ref{prop11} in Section \ref{eq}, Lemmas \ref{lem4} to \ref{lem:rightWall}, 
Propositions \ref{prop:2} to \ref{th2},
Corollaries \ref{constraint} and \ref{c2}, Theorem \ref{th2_pts} in Section \ref{sec:fluc}, 
Section \ref{sec:stability} on stability analysis, 
Section \ref{Rt} on revisiting the examples,
Section \ref{sec:basin} on approximating the basin of attraction, 
Section \ref{sec:concludingRemarks} on concluding remarks, and
the Appendix including Lemma \ref{lem2}
and the proofs of and further elaborations on all of the results.
}
    \thanks{This project was partly funded by Alberta Environments and Parks.}
}

\begin{document}
\maketitle

\begin{abstract}
Decision-making individuals often imitate their highest-earning fellows rather than optimize their own utilities, due to bounded rationality and incomplete information.
%on how their utilities depend on their own and others' decisions.
Perpetual fluctuations between decisions have been reported as the dominant asymptotic outcome of imitative behaviors, yet little attempt has been made to characterize them, particularly in heterogeneous populations. 
We study a finite well-mixed heterogeneous population of individuals choosing between the two strategies, cooperation and defection, and earning based on their payoff matrices that can be unique to each individual.
At each time step, an arbitrary individual becomes active to update her decision by imitating the highest earner in the population. 
We show that almost surely the dynamics reach either an equilibrium state or a minimal positively invariant set, a \emph{fluctuation set}, in the long run.
In addition to finding all equilibria, for the first time, we characterize the fluctuation sets, provide necessary and sufficient conditions for their existence, and approximate their basins of attraction. 
%In particular, given the distribution of the individuals over the payoff matrices, we explicitly identify all possible asymptotic outcomes for the total number of cooperators in the population. 
We also find that exclusive populations of individuals playing coordination or prisoner's dilemma games always equilibrate, implying that cycles and non-convergence in imitative populations are due to individuals playing anticoordination games. 
Moreover, we show that except for the two extreme equilibria where all individuals play the same strategy, almost all other equilibria are unstable as long as the population is heterogeneous. 
Our results theoretically explain earlier reported simulation results and shed new light on the boundedly rational nature of imitation behaviors.
%and pave the way for future research on controlling such populations. 
\end{abstract}


% \fi
\IEEEpeerreviewmaketitle

\section{Introduction}
Imitation is a commonly adopted decision-making strategy, especially under bounded rationality \cite{seel2011encyclopedia,apesteguia2007imitation,shea2009imitation}. 
Animals learn skills \cite{byrne2003imitation} and find valuable food \cite{laland2004social}, managers make profitable investments \cite{scharfstein1990herd}, firms avoid falling behind their rivals \cite{posen2013power,lieberman2006firms}, and humans vaccinate against infectious diseases \cite{oraby2014influence,bauch2005imitation,fu2010imitation}--all by mimicking successful others.
Many seemingly irrelevant processes such as death-birth can also be modeled by imitation: a dead cell can be thought of as adapting to the cell that divides to occupy the vacancy left by the dead \cite{seel2011encyclopedia}.
In the context of \emph{evolutionary game theory}, the recently appealing field that applies game theory to the study and control of spreading behaviors in evolving populations \cite{riehl2016towards,ramazi2020global, madeo2014game,zhu2016evolutionary,ramazi2018homophily}, imitation is often modeled by the \emph{imitation update rule}, leading to the \emph{imitation dynamics} \cite{tan2016analysis}, \cite{como2020imitation}, \cite{zino2017imitation} and its mean field approximation, the famous replicator dynamics \cite{gao2020passivity, abbass2015n}.
Compared to \emph{myopic best-response} \cite{ramazi2017asynchronous,ramazi2020convergence}, where individuals choose the action maximizing their current utility, imitation more often leads to selfish behavior in certain human experiments \cite{van2015focus}.
This higher selfishness, however, does not lead to higher satisfaction.
Although some continuous imitation dynamics equilibrate in unstructured populations \cite{barreiro2018constrained},  
fewer networks of imitators are guaranteed to reach an equilibrium state, where all individuals are satisfied with their actions \cite{ramazi2016networks, riehl2018survey}.
Therefore, unlike best-responders, non-equilibration and long-term oscillations are more observed among imitators \cite{henderson2016alternative, cimini2015dynamics, govaert2017convergence}, possibly due to their limited rationality \cite{vilone2012social}.
Although representing dissatisfaction in social groups and stock markets, the oscillations may also indicate life in organisms or mobility in research organizations, emphasizing their key role in different fields.  
The formulation of and conditions for these oscillations, however, remains an open yet important problem. 

%Therefore, we study imitation dynamics to better understand decision-making behaviors with limited rationality. 

% repeat this in the next paragraph
%By means of numerical simulations, it was shown in that even for homogeneous agents updating synchronously, imitation does not generally result in equilibrium convergence, although it appears likely for coordination games. \cite{cimini2015dynamics}

%In the context of \emph{Evolutionary game theory}, the application of game theory to study the spread of behaviors in evolving populations \cite{nowak2006evolutionary}, ``Imitation dynamics'' is one of the most widely used models to study the imitative behaviors mentioned above \cite{hofbauer2003evolutionary}-\cite{traulsen2009exploration}. 

% 
%Two of the most commonly adopted imitation rules are \emph{unconditional imitation} and \emph{replicator} \cite{cheng2016decomposed,mei2017dynamic}. 
%Under unconditional imitation, an individual observes her neighbors' (and in some settings, her own) payoffs and imitates the one with the highest payoff. 
%Under replicator, an individual randomly selects a neighbor and decides to imitate her with the probability proportional to the payoff difference if the neighbor has a higher payoff \cite{xia2012role}. 
%Numerical simulations have shown that in certain structured population, the long-term behavior of the two can be greatly different \cite{li2012cooperative}. 
Due to its extreme boundedly rational nature and broad application, we focus on the unconditional imitation rule \cite{xia2012role,riehl2017controlling}, where an individual observes her neighbors' (and in some settings, her own) payoffs and imitates the one with the highest payoff.
%We focus on the unconditional imitation rule due to its extreme boundedly rational nature and broad application \cite{cheng2016decomposed, riehl2017controlling} and 
We also consider the widely applied binary choices of strategies \emph{cooperation} and \emph{defection} \cite{madeo2014game, tan2016analysis, ramazi2016networks}. 
Most existing results on the (unconditional) imitation rule are empirical.
While cooperation is reported to hardly sustain under imitation, simulation results show that it can be promoted by the introduction of moderate random exploration \cite{szolnoki2014defection, traulsen2009exploration, fudenberg2008monotone}, increase of heterogeneity \cite{santos2006evolutionary}, or combination with some other dynamics \cite{roca2009imperfect}. 
Furthermore, researchers have shown that chaotically changing spatial patterns can be generated in structured populations \cite{nowak1992evolutionary, szolnoki2014defection} and the long-term behaviors vary by initial states with different fractions of cooperative agents \cite{piter2016, li2012cooperative}. 
It remains concealed whether the chaotic fluctuations can also happen in well-mixed populations, and if so, under what conditions. 
The importance of well-mixed populations is upheld by real-world cases such as stock markets and academic institutes, where individuals observe the highest earners' actions and payoffs. 
On the other hand, many theoretical studies related to this topic focus on homogeneous populations \cite{xu2017cooperation, cheng2015modeling, dyer2002convergence}. 
Different approaches have been used to show that cooperation can never emerge under imitation dynamics in a well-mixed homogeneous population where each individual performs a continuously mixed strategy \cite{tan2016analysis}, and convergence analysis has been performed for networks of coordinating agents \cite{riehl2018incentive} and well-mixed populations of agents playing public goods games \cite{govaert2017convergence}. 
Less attention has been paid to heterogeneous populations, where agents earn differently, even though slight changes of the agents' payoffs can remarkably promote cooperation in imitation dynamics \cite{su2019evolutionary}.
This is despite the literature pointing out the limitation of structured homogeneous populations in understanding human behavior, and more attention is demanded on heterogeneous populations \cite{grujic2014comparative}. 

We consider a finite, well-mixed, heterogeneous population of interacting agents playing against each other, choosing cooperation or defection, and earning according to their possibly unique payoff matrices.
The agents become active asynchronously to update their strategies according to the (unconditional) imitation update rule. 
We take the distribution of cooperators over the payoff matrices as the state of the system and investigate its long-term behavior.
The solution trajectory either enters a non-singleton minimal positively invariant set or reaches an equilibrium. 
By developing a graphical representation of the dynamics, for the first time, we characterize all these positively invariant sets for a given initial condition and approximate their basins of attraction. 
Moreover, we determine the equilibria of the dynamics and show that--except for the two extreme equilibria where all agents play the same strategy--no equilibrium is asymptotically stable as long as the population is heterogeneous.
In addition to these general results, we investigate the special cases where the payoff matrices, although different, are all of the form of one of the well-known $2\times2$ games.
In line with previous simulation results, we find the dynamics equilibrate when the agents all play prisoner's dilemma or  coordination games, but may fluctuate when some play anticoordination games (such as snowdrift), highlighting the key role of individuals with anticoordination payoff matrices in forming oscillations.
Our contribution is fourfold: 
\emph{i)} We show that heterogeneity is enough to cause oscillations even in a well-mixed population. 
Reported fluctuations are, thus, not necessarily due to how agents are connected. 
\emph{ii)} We explicitly characterize the oscillations and discover that they are caused by the existence of individuals who highly value the action opposite to the one taken by the majority, and hence, agents with a base game of anticoordination are the key components.
\emph{iii)} Results on stability shed new lights on the bounded rational nature of imitation behaviors, pleading ideas on controlling such populations for future studies, such as changing the number of certain individuals, forcing a single agent to change her strategy to break an undesired equilibrium state.
\emph{iv)} The graphical representation method has the potential to be applied to other population dynamics to provide intuitive insight on the asymptotic behavior. 

%\par
%The rest of the paper is organized as follows. In Section II, we present the framework, formulate the problem and provide an example. In Section III, we provide a graphical representing method of the population dynamics. In Section IV, we find the equilibrium states of the population dynamics, which is followed by the identification of the `fluctuation sets', sets of states that the population keeps changing between without ending.In Section VI, we proceed to the stability analysis of the equilibrium states. In Section VII, we give a solution to the example provided in Section II, and based on the solution we conclude the conditions under which the population ends up in a fluctuation set. Then in Section VIII, we proceed to the convergence analysis of different types of population and finish the paper in Section IX with the discussion and concluding remarks.

\section{Decision-making under imitation updates}
We consider a well-mixed population of $n$ agents, labeled by $1,\ldots, n$, that are participating in a population game evolving over time $t=0,1\ldots$. 
Each agent can choose between one of two strategies, \emph{cooperation} $(C)$ and \emph{defection} $(D)$.
At each time step $t$, every agent earns an accumulated payoff against the population, and one random agent becomes active to mimic the strategy of the most successful agent in the population. 
The four possible payoffs of an agent $i$
against agent $j$, where $i,j\in\{1,\ldots,n\}$, are summarized by the $2\times 2$ payoff matrix
\begin{equation*}
	\pi^i =
		\bordermatrix{
		    &  C &D \cr 
		 C  & \textcolor{Blue}{R_i} & \textcolor{Emerald}{S_i} \cr
		 D  & \textcolor{RawSienna}{T_i} & \textcolor{Red}{P_i} }, \qquad
		   \textcolor{Emerald}{S_i},\textcolor{Blue}{R_i}, \textcolor{RawSienna}{T_i}, \textcolor{Red}{P_i} \in {\rm I\!R}\\
\end{equation*}
whose $lk^\text{th}$ entry, $l,k\in\{C,D\}$, corresponds to the strategy pair $l$-against-$k$, where agent $i$ plays strategy $l$ and agent $j$ plays strategy $k$.
Let $s^i$ denote the strategy of agent $i$ in the vector form, which is either  
$
s^C \dfb
\begin{bmatrix}
	1 & 0 
\end{bmatrix}^\top
$
for cooperation or 
$
s^D \dfb
\begin{bmatrix}
	0 & 1 
\end{bmatrix}^\top
$
for defection. 
Then the accumulated payoff or the \emph{utility} of agent $i$ at time $t$ against the population is
\begin{equation*}
	u^i(t) = s^i(t)^\top \pi^i\,  
	    \begin{bmatrix}
	        n^C(t) & n-n^C(t)
        \end{bmatrix}^\top,
\end{equation*}
%where $s^C(t) \dfb 
%\begin{bmatrix}
%	n^C(t) & n-n^C(t)
%\end{bmatrix}^\top$
where $n^C(t)$ denotes the number of cooperators in the whole population at time $t$.
The \emph{imitation update rule} for agent $i$ active at time $t$ dictates that 
she revises her strategy at time $t+1$ to that of the agent 
with the maximum utility in the population at time $t$. 
In case, two agents with different strategies, i.e., a cooperator and a defector, both earn the highest utility, agent $i$ sticks to 
her current strategy. 
Equivalently, agent $i$ switches her strategy if and only if her strategy is not played by a highest-earning agent:
\begin{equation}		\label{updateRule}
	s^i(t+1) =
	\begin{cases}
		s^i(t)								& s^i(t) \in \str M(t)		\\
		\mathbf{1}-s^i(t)	&  s^i(t) \not\in \str M(t)
	\end{cases},
\end{equation}
where $\mathbf{1} = 
\begin{bmatrix}
	1 & 1 
\end{bmatrix}^\top$ and $\str M(t)$ is the set of the strategies of agents with maximum utility at time $t$, i.e., 
\begin{equation*}
	\str M(t) \dfb \left\{s^j(t)\,\Big|\, j\in\arg\max_{k \in \{1,\ldots,n\}} u^k(t) \right\}.
\end{equation*}

The agents can be thought of as firms competing over the markets of a product.
They produce the product either with ($C$) or without ($D$) some features. 
Payoffs are then their seasonal shares of the market. In each season, a firm may imitate the one with the largest share of the market in the last season to add/delete features to/from its own product.  
%In a different scenario, people involved in a pyramid scheme can be regarded as agents who decide either to continue participating in the scheme ($C$) or withdraw ($D$). 
%Occasionally, they may revise their decisions by blindly following the highest earning individual. 
As another example, graduate students or scholars often decide to work on the scientific topic that their successful top-ranked fellows do, despite the differences in their capabilities.
The model can also be applied to scenarios where agents do not actually imitate, but their behavior are interpreted as ``imitation.''
For example, cells in a cancerous organ are either normal ($C$) or cancer cells ($D$). 
All cells share the same glucose in the extracellular fluid, and the utility for each agent is its ``reproduction capability'', determined by its cost of glucose in division and the amount of glucose it receives.
When a cell dies, the vacant left by is filled by the offspring of the fittest neighbor, either a normal or cancer cell, which can be viewed as the dead cell adapting to the neighbor cell with the highest utility.

However, is it rational for an agent to imitate the strategy of another agent with a different utility? 
Regardless of the answer, such behavior is observed in reality \cite{scharfstein1990herd,seel2011encyclopedia,laland2004social}.
The revenue that different firms earn from producing the same product is not the same, neither are the quality and quantity of the papers that different scholars publish on the same topic, 
and the same goes with the glucose consumption of normal and cancer cells.
Nevertheless, in all of these cases, some individuals do imitate.
The fact that such decision-making may seem irrational, explains exactly why imitation is often not considered a (myopically) rational choice \cite{govaert2019relative}. 

%To illustrate·, the agents can be considered as cells in a cancerous organ, who can be either normal cells ($C$) or cancer cells ($D$). At each update, a cell is chosen randomly for death, the vacant site is then filled by the offspring of either normal cells or cancer cells, depending on their abilities of giving birth (payoff).

We categorize all agents with the same payoff matrix into the same \emph{type} and assume that there are altogether $m, 0<m\leq n$, types of agents.
We label the types by $1,\ldots,m$ according to the descending order of the number of agents in the same type.
The heterogeneity of the population is then characterized by 
\begin{equation*}
	p \dfb (n_1,n_2,\ldots,n_m),
\end{equation*}
where $n_i$ is number of type-$i$ agents and $n_i\geq n_j$ if $i>j$. 
We take the distribution of cooperators over the types as the state of the system: 
\begin{equation*}
	x(t) \dfb
	\left(
		x_1(t), x_2(t),  \ldots, x_m(t)
	\right),
\end{equation*}
where $x_j\in\Z_{\geq0}$ is the number of cooperators of type $j$.
The state space will then be
\begin{equation*}
    \mathcal{X} \dfb
    \left\{ x\in \Z_{\geq0}^m \,\Big|\, 
    x_j\leq n_j ~\forall j \in \{1,\ldots,m\} \right\}.
\end{equation*}
We denote the total number of cooperators at a state $x$ by $n^C(x)$ and often simplify the notation $n^C(x(t))$ to $n^C(t)$ for a given time $t$.
Equation \eqref{updateRule} together with the agents' activation mechanism govern the dynamics of $x(t)$, which we refer to as the \emph{(asynchronous) imitation dynamics}.   
We assume that at each time, every agent has the chance to update; however, only one becomes active, making the dynamics asynchronous. 
For example, it might be the case that every agent has a Poisson clock--not synced with others--that upon ticking, she becomes active to update her strategy. 
Then each time $t$ will be when an agent's clock ticks. 
We refer to the order of the active agents over time as the \emph{activation sequence}, which is generated from a random process. 
The randomness of the activation sequence, implies that any finite sequence of agents shows up in the activation sequence \emph{almost surely}, i.e., with probability 1 as time goes to infinity.
Our goal is to study the evolution of $x(t)$ and $n^C(t)$
under the imitation dynamics, particularly, their asymptotic behaviors. 
%We assume the activations sequence is random.
%By randomness we mean the assumption that the activation sequence is \emph{persistent}, which can be stated as follows.
%\begin{assumption}
%	Every agent activates infinitely many times as time goes to infinity.
%\end{assumption}
%In the next section, we show that $x(t)$ either converges to a single state or fluctuates between several states close to each other. 

\section{Example} \label{sec:example}
\begin{figure*}%[t!]
    \centering
    \begin{subfigure}[h]{0.5\textwidth}
        \centering
        \captionsetup{width=0.8\linewidth, font=footnotesize}
        \includegraphics[width=\linewidth,height=\textheight,keepaspectratio]{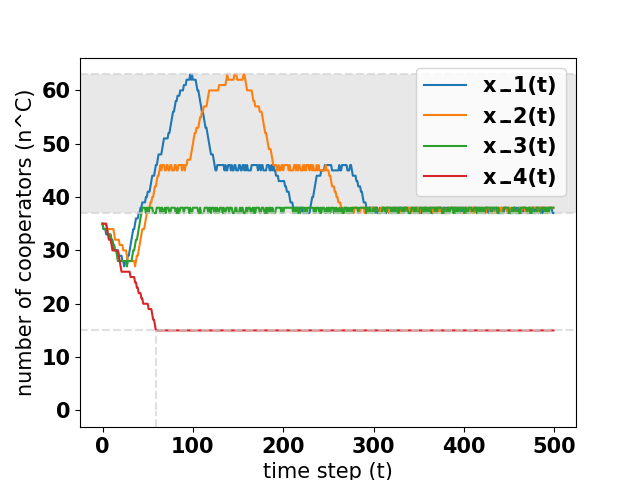}
        \caption{}
    \end{subfigure}%
    \begin{subfigure}[h]{0.5\textwidth}
        \centering
        \captionsetup{width=0.8\linewidth, font=footnotesize}
        \includegraphics[width=\linewidth,height=\textheight,keepaspectratio]{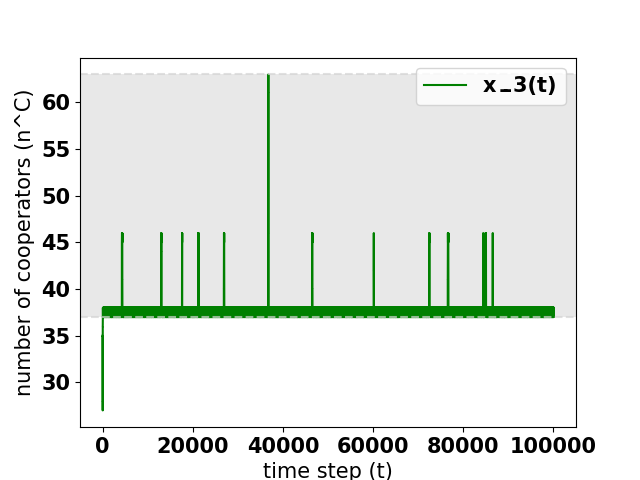}
        \caption{}
    \end{subfigure}
    \caption{
	\textbf{The evolution of $n^C(x)$ in Section \ref{sec:example} under four different activation sequences.} 
	\textbf{a)} Trajectories $x^1(t)$, $x^2(t)$, $x^3(t)$, $x^4(t)$ all have the same initial condition $x(0) = (13,18,4,0)$ with $n^C(0) = 35$.
	However, under different activation sequences, $x^4(t)$ reaches an equilibrium at $t=59$ and stays there afterwards, while $x^1(t)$ and $x^2(t)$ keep fluctuating between those states with corresponding $n^C$ values between $n^C = 37$ and $n^C = 63$, and $x^3(t)$ keeps fluctuating between those states with corresponding $n^C$ values between $37$ and $38$. 
	\textbf{b)}
	The evolution of trajectory $x^3(t)$ over $100,000$ time steps.
	The trajectory keeps fluctuating between those states with corresponding $n^C$ values between $n^C = 37$ and $n^C = 63$. 
	However, it fluctuates between $37$ and $38$ most of the time.}
	\label{fig1}
\end{figure*}

%\begin{example} \label{ex1}
To better understand the dynamics, we provide the following example, where both equilibration and perpetual cycling can happen for the same initial condition.
Consider a population comprising four types of agents with the type distribution vector $p = (13,18,5,64)$ and payoff matrices
\begin{align*}
     \pi_1 &= \left( \begin{array}{cc} 
	\scalebox{0.8}{-0.3015} & \scalebox{0.8}{0.3685} \\
	\scalebox{0.8}{1.2} & \scalebox{0.8}{-0.3}
	\end{array} \right)\text{, }
	\pi_2 = \left( \begin{array}{cc}
		\scalebox{0.8}{1.68} & \scalebox{0.8}{-5.12} \\
		\scalebox{0.8}{0.7} & \scalebox{0.8}{-1.3}
		\end{array} \right)\text{, }\\
	\pi_3 &= \left( \begin{array}{cc}
		\scalebox{0.8}{0.1} & \scalebox{0.8}{0.5} \\
	\scalebox{0.8}{1.0976} & \scalebox{0.8}{0.1216} 
		\end{array} \right)\text{, }
		\pi_4 = \left( \begin{array}{cc} 
			\scalebox{0.8}{-0.29365} & \scalebox{0.8}{0.95635} \\
			\scalebox{0.8}{1.185625} & \scalebox{0.8}{-1.309375}
			\end{array} \right)\text{, }
\end{align*}
where $\pi_i, i=1,\ldots,4$, denotes the payoff matrix of type-$i$ agents.
Note the use of a lower index for type quantities, e.g., $\pi_i$, versus the use of an upper index for agent quantities, e.g., $\pi^i$.
Payoff matrices $\pi_1$, $\pi_2$, and $\pi_3$ are those of snowdrift games and 
$\pi_4$ is a payoff matrix of a coordination game. 
Under the imitation update rule (\ref{updateRule}),
starting from the same initial condition $x(0) = (13,18,4,0)$ with $n^C(x(0)) = 35$, based on the activation sequence, the solution trajectory $x(t)$ may either reach an equilibrium state or end up fluctuating between several states with the total number of cooperators ($n^C$) in the interval $(37, 63)$ (Fig.~\ref{fig1}).
We will return to this example in Section \ref{Rt}. 
% \begin{figure}
% 	\centering
% 	\begin{subfigure}%[t]{0.5\textwidth}
%         \centering
%         \includegraphics[width=\linewidth,height=\textheight,keepaspectratio]{Figure_x1-4.png}
%         \caption{Lorem ipsum}
%     \end{subfigure}%
%     \begin{subfigure}%[t]{0.5\textwidth}
%         \centering
%         \includegraphics[width=\linewidth,height=\textheight,keepaspectratio]{Figure_x5.png}
%         \caption{Lorem ipsum}
%     \end{subfigure}%
% 	%\captionsetup{font=tiny}
% 	\caption{
% 	\textbf{The evolution of $n^C(x)$ in Section \ref{sec:example} under five different activation sequences.}
% 	Trajectories $x^1(t)$, $x^2(t)$, $x^3(t)$, $x^4(t)$, and $x^5(t)$, all have the same initial condition $x(0) = (13,18,4,0)$ with $n^C(0) = 35$.
% 	However, under different activation sequences, $x^4(t)$ reaches an equilibrium and stays there afterwards, while others keep fluctuating between those states with corresponding $n^C$ values between $n^C = 37$ and $n^C = 63$.}
% 	\label{fig1}
% \end{figure}
%\end{example}

%%%%%%%%%%%%%%%%%%%
\section{Graphical representation of imitation dynamics}
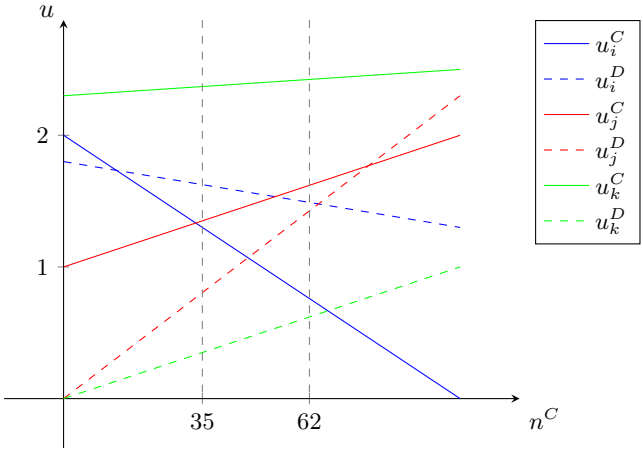
\begin{figure}\label{x1}
	\centering
	\begin{tikzpicture}
		\begin{axis}[
			standard,
			xlabel=$n^C$,
			xtick={35,62},
			ylabel=$u$,
			%xtick={0,1.7},
			%xticklabels={0,$t_1$},
			%ytick={0,21},
			%yticklabels={0,$v_0$}
			%scaled ticks=base 10:-11
			%axis lines=middle,
			xmin=0, xmax=100,
			legend pos= outer north east,
			legend style={font=\small},
			every axis y label/.style={
            at={(ticklabel* cs:1.02)},
    		anchor=east,
}
		]
		%\coordinate (A) at (0,2);
		%\coordinate (B) at (100,0);
		%\draw [name path=A--B] (A) -- (B);
		%\coordinate (C) at (0,1.8);
		%\coordinate (D) at (100,1.3);
		%\draw [name path=C--D] (C) -- (D);
		%\path [name intersections={of=A--B and C--D,by=E}];
		%\node [fill=red,inner sep=1pt,label=-90:$m$] at (E) {};

		\addplot [
			name path global=funone,
			domain=0:100, 
			samples=10, 
			color=blue,
			] coordinates
			{(0,2)
			(100,0)};
		\addlegendentry{$u_i^C$}

		\addplot [
			name path global=funtwo,
			domain=0:100, 
			samples=10, 
			color=blue,
			style=dashed
			] coordinates
			{(0,1.8)
			(100,1.3)};
		\addlegendentry{$u_i^D$}
		%Here the blue parabloa is defined
		\path [name intersections={of=funone and funtwo,by=E}];
		%\node [circle,fill=black,inner sep=1.5pt,label=90:$m$] at (E) {};

		\addplot [
			domain=0:100, 
			samples=10, 
			color=red,
			] coordinates
			{(0,1)
			(100,2)};
		\addlegendentry{$u_j^C$}

		\addplot [
			domain=0:100, 
			samples=10, 
			color=red,
			style=dashed
			] coordinates
			{(0,0)
			(100,2.3)};
		\addlegendentry{$u_j^D$}
		%Here the blue parabloa is defined
		\addplot [
			domain=0:100, 
			samples=10, 
			color=green,
			] coordinates
			{(0,2.3)
			(100,2.5)};
		\addlegendentry{$u_k^C$}

		\addplot [
			domain=0:100, 
			samples=10, 
			color=green,
			style=dashed
			] coordinates
			{(0,0)
			(100,1)};
		\addlegendentry{$u_k^D$}

		\draw [black, fill] (35 ,162) circle (2pt) node [above] {$m$};
		\draw [gray, dashed] (35,162) -- (35,0);
		\draw [black, fill] (62 ,162) circle (2pt) node [above] {$n$};
		\draw [gray, dashed] (62,162) -- (62,0);

		%\addplot[thick,color=black] coordinates { (0,21) (1.7,21) (8.7,0) };

		\end{axis}
	\end{tikzpicture}
	%\captionsetup{font=tiny}
	\caption{
	\textbf{Graphical representation of the imitation dynamics.} 
	Cooperative and defective lines are represented by solid and dashed lines, respectively. 
% 	Line \textcolor{black}{$u_k^C$} is the max line at every $n^C$. 
% 	Assume that at some time $T$, 
% 	$n^C(T)=35$,
% 	\textcolor{black}{$u_k^C$} is inactive, and
% 	\textcolor{black}{$u_i^D$} is active. 
% 	If the active agent at time $T$ is a cooperator, then she will change her strategy to defection, and we need to move one unit left on the $n^C$-axis to obtain $n^C(T+1)=34$.
% 	Assume, instead, that $n^C(T)=62$,
% 	\textcolor{black}{$u_k^C$} is inactive, and
% 	\textcolor{black}{$u_j^C$} is active. 
% 	Now if the active agent at time $T$ is a defector, then she will switch to cooperation, and we need to move one unit right on the $n^C$-axis to obtain $n^C(T+1)=63$.
	} 
	\label{fig2}
\end{figure}

We provide a graphical representation of how the agents' strategies evolve over time, which turns out to be the key in characterizing the oscillations and illustrating the convergence results.  
The utility of every type-$i$ agent who chooses to cooperate 
at time $t$, denoted by $u_i^C(t)$, 
and the utility of every type-$i$ agent who chooses to defect
at time $t$, denoted by $u_i^D(t)$, are calculated by
% \begin{equation*}		
% 	\begin{cases}
% 		u_i^C(t) =  \textcolor{Blue}{R_i} n^C(t) + \textcolor{Emerald}{S_i} \left(n-n^C(t)\right) \text{,} \\
% 		u_i^D(t) =	\textcolor{RawSienna}{T_i} n^C(t) + \textcolor{Red}{P_i}\left(n-n^C(t)\right) \text{.}
% 	\end{cases}
% \end{equation*} 
% By rearranging the above expressions we get
\begin{equation*}	
    % \label{s_expr}	
	\begin{cases}
		u_i^C(t) = 	\underbrace{(\textcolor{Blue}{R_i}-\textcolor{Emerald}{S_i})}_{a_i}n^C(t) + \underbrace{\textcolor{Emerald}{S_i}n}_{b_i} \text{,} \\
		u_i^D(t) = 	\underbrace{(\textcolor{RawSienna}{T_i}-\textcolor{Red}{P_i})}_{c_i}n^C(t) + \underbrace{\textcolor{Red}{P_i}n}_{d_i}	\text{.}
	\end{cases}
\end{equation*}
Since $a_i,b_i,c_i$ and $d_i$ are constants, 
the utility of type-$i$ agents is linear 
in $n^C$. 
So as time evolves, 
$u_i^C(t)$ and $u_i^D(t)$ move along 
the following two lines in the plane spanned by the two axes $n^C$ (the $x$-axis) 
and utility, denoted by $u$, (the $y$-axis):
\begin{gather}  \label{eq3}
	u_i^C\left(n^C\right) = a_i n^C + b_i,\qquad
	u_i^D\left(n^C\right) = c_i n^C + d_i .
\end{gather}
We call $u_i^C$ and $u_i^D$ the \emph{cooperative} and \emph{defective lines} of type-$i$ agents.
Given a state $x\in\X$, if $x_i>0$, we say that the $u_i^C$ line is \emph{active (by $x$)} since at least one type-$i$ agent is cooperating; otherwise, the $u_i^C$ line is \emph{inactive (by $x$)}. 
Similarly, if $x_i<n_i$, we say that the $u_i^D$ line is \emph{active (by $x$)} since at least one type-$i$ agent is defecting; otherwise, the $u_i^D$ line is \emph{inactive (by $x$)}.
For example, in Fig.~\ref{fig2}, assume that for some time $T$, $x_k(T)=0$ and $x_i(T)<n_i$. 
Then $u_k^C$ is inactive but $u_i^D$ is active by $x(T)$.
In total there are $2m$ lines in the plane, 
at least $m$ of which are active at each time.
We call the line(s) with the maximum $u$ value at $n^C$, the \emph{max line(s) at $n^C$}
(e.g., $u_k^C$ in Fig.~\ref{fig2} is the max line at every $n^C$)
\footnote{There may be more than one max line at $n^C$; however, for sake of simplicity, we often say \emph{the max line at $n^C$ is cooperative (defective)} instead of \emph{all the max lines at $n^C$ are cooperative (defective)}.}.

Among the lines active by $x\in\X$, we often refer to the max at $n^C$ as the \emph{max active-by-$x$ line at $n^C$} or simply \emph{max active line at $n^C$} if it is clear from the context.
Back to our example regarding Fig.~\ref{fig2}, if we additionally assume that $n^C(T) = 35$, then the max active line at $35$ is $u_i^D$. 
If the active agent at time $T$ is a cooperator, then she will change her strategy to defection, and we need to move one unit left on the $n^C$-axis to obtain $n^C(T+1)=34$.
Assume, instead, that $n^C(T)=62$,
$u_k^C$ is inactive, and $u_j^C$ is active. 
Now if the active agent at time $T$ is a defector, then she will switch to cooperation, and we need to move one unit right on the $n^C$-axis to obtain $n^C(T+1)=63$.
In general, if the active agent switches her strategy to cooperation at $t+1$, then
$n^C(t+1)=n^C(t)+1$. If the active agent switches to defection, then
$n^C(t+1)=n^C(t)-1$. 
Otherwise, we will have $n^C(t+1)=n^C(t)$ and
the same active-line configuration for $t+1$ as that at time $t$.
Hence, in order to find $n^C(t+1)$ from $n^C(t)$, we need to move one unit
right on the $n^C$-axis if the max active lines at $n^C(t)$ are all cooperative, 
and move to left if the max active lines at $n^C(t)$ are all defective, and not move otherwise (Fig.~\ref{fig2}). 
In this way, we can monitor the evolution over time of $n^C(t)$, but only partly that of $x(t)$ 
since different values of $x$ may correspond to the same value of $n^C$.

\section{Equilibrium states}\label{eq}
A state $x^*$ is an equilibrium of the imitation dynamics if whenever the state $x(t)$ equals $x^*$ it remains there afterwards, regardless of the activation sequence. 
This means that no agent would switch her strategy, which in view of the imitation update rule, implies that every agent in the population has chosen a strategy that is played by one with the highest payoff. 
This condition is clearly satisfied if all agents are cooperating or all agents are defecting. 
Indeed, there will be no other agent with a different strategy to mimic in this case.
This results in the two equilibria $x=(0,\ldots,0)$ and $x=(n_1,\ldots,n_m)$ which we refer to as the \emph{extreme equilibrium states}.
The other possibility for a state to be an equilibrium is the existence of at least one cooperator and one defector who both have the highest payoff in the population. Namely, they earn the same payoff, and no other agent earns more. 
Let $\mathcal{X}^*\subseteq\X$ denote the set of all equilibrium states.
\begin{theorem} \label{th1}
    Given a state $x\in\mathcal{X}$, it holds that 
	$x \in \mathcal{X}^*$ if and only if exactly one of the following is true:
	\begin{enumerate}
		\item $x=(0,\ldots,0)$ or $x=(n_1,\ldots,n_m)$;
		\item there exist two (not necessarily different) agent types $i$ and $j$, 
		such that $x_i \neq 0$ and $x_j\neq n_j$, 
		$(a_i-c_j)n^C(x) = d_j-b_i$, and for every other type $k$, it holds that
			\begin{align*}
			\begin{cases}
				x_k=0 & \text{if } a_kn^C(x)+b_k>a_in^C(x)+b_i \text{,} \\
				x_k=n_k & \text{if } c_kn^C(x)+d_k>c_jn^C(x)+d_j \text{.}	
			\end{cases}
			\end{align*}
	\end{enumerate}
\end{theorem}
%\bigskip

Theorem \ref{th1} enables us to systematically loop over all states to find the equilibria: for each $n^C\in\{0,\ldots,n\}$, we check the conditions in Theorem \ref{th1} for all states $x\in\X$ whose number of cooperators equals $n^C$.
When checking for possible $k$s in the theorem, one may use the following property that is straightforward to show. 
For each type-$k$ agent, we have
\begin{equation} \label{property}
	\begin{cases}
		x_k\neq n_k &\text{if } n_k>n^C(x) \text{,} \\
		x_k\neq 0 & \text{if } n_k>n-n^C(x) \text{.}	
	\end{cases}
\end{equation}

Back to the graphical representation, the corresponding representation of $x^*\in\X^*$ must be a point such that 
$n^C$ neither moves right nor left from $n^C(x^*)$ over time. 
In addition to the two trivial cases of $n^C = 0$ and $n^C = n$, 
this happens only when a cooperative and a defective max active line at $n^C(x^*)$ intersect.
\begin{figure}
	\centering
	\begin{tikzpicture}
		\begin{axis}[
			standard,
			xlabel=$n^C$,
			xtick={20,37.5,80},
			ylabel=$u$,
			legend pos=outer north east,
			legend style={font=\small},
			every axis y label/.style={
    		at={(ticklabel* cs:1.02)},
    		anchor=east,
}
		]
		\addplot [
			name path=P1,
			domain=0:100, 
			samples=10, 
			color=blue,
			]coordinates
			{(0,0)
			(100,5)};
		\addlegendentry{$u^C_i$}
		
		\addplot [
			name path=P2,
			domain=0:100, 
			samples=10, 
			color=blue,
			style=dashed
			]coordinates
			{(0,0.75)
			(100,2)};
		\addlegendentry{$u^D_i$}

		]
		\addplot [
			name path=P3,
			domain=0:100, 
			samples=10, 
			color=red,
			]coordinates
			{(0,4)
			(100,1.1875)};
		\addlegendentry{$u^C_j$}
		
		\addplot [
			name path=P4,
			domain=0:100, 
			samples=10, 
			color=red,
			style=dashed
			]coordinates
			{(0,3)
			(100,0)};
		\addlegendentry{$u^D_j$}

		\path [name intersections={of=P1 and P2,by=A}];
		\node [circle,fill=black,inner sep=1.5pt,label=90:$a$] at (A) {};
		\path [name intersections={of=P1 and P4,by=B}];
		\node [circle,fill=black,inner sep=1.5pt,label=90:$b$] at (B) {};
		\path [name intersections={of=P1 and P3,by=C}];
		\node [circle,fill=black,inner sep=1.5pt,label=90:$c$] at (C) {};
		\path [name intersections={of=P2 and P4,by=D}];
		\node [circle,fill=black,inner sep=1.5pt,label=90:$d$] at (D) {};
		\path [name intersections={of=P3 and P2,by=E}];
		\node [circle,fill=black,inner sep=1.5pt,label=90:$e$] at (E) {};
		%Here the blue parabloa is defined
		%\draw [black, fill] (16.5,82) circle (2pt) node [right] {$a$};
		%\fill[black,name intersections={of=P1 and P2,total=\t}] ;
		%\path [name intersections={of=P1 and P2,by={CS}}];
		%	\foreach \s in {1,...,\t}{(intersection-\s) circle (2pt) node {\footnotesize\s}};
		%\node[red,scale=3] at (intersection of  line 1 and line 2){.};
		
		%\draw [black, fill] (50.5 ,148) circle (2pt) node [right] {$b$};
		%\draw [black, fill] (70 ,190) circle (2pt) node [right] {$c$};
		%\node[dot=E] (E) at (CS) {};

		\draw [gray, dashed] (A) -- (20,0);
		\draw [gray, dashed] (B) -- (37.5,0);
		\draw [gray, dashed] (E) -- (80,0);

		\end{axis}
	\end{tikzpicture}
	\caption{\textbf{Intersection points.} 
	Points $a$, $b$, and $e$ are intersection points, but points $c$ and $d$ are not. 
	Among these intersection points, equilibria can be reached only at $n^C(e)$ since $b$ is non-integer and $a$ cannot be intersected by max active lines as both $u^C_j$ and $u^D_j$ are above $a$ at $n^C(a)$ and at least one of them is active at any time.
	Any equilibrium with intersection point $e$ is potentially attracting since $u_j^C(79) > u_i^D(79)$.} 
	\label{fig3}
\end{figure}
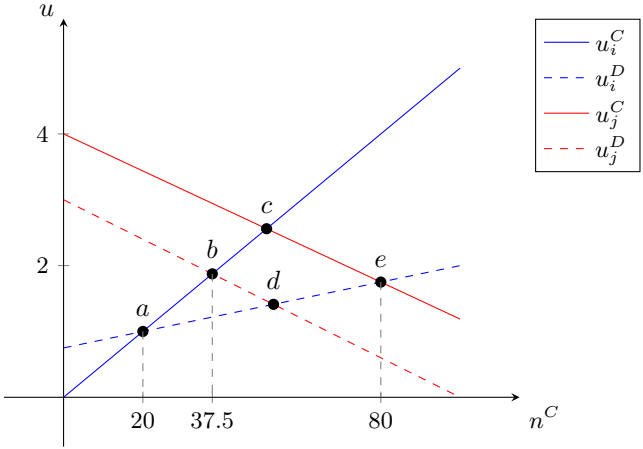
\begin{proposition} \label{prop11}
	$x^* \in \mathcal{X}^*$ if and only if exactly one of the following is true:
	\begin{enumerate}
	    \item $n^C(x^*) = 0$ or $n^C(x^*) = n$;
		\item there exist a cooperative max and a defective max line at $n^C(x^*)$, both active by $x^*$, which intersect at $n^C(x^*)$.
	\end{enumerate}
\end{proposition}

Given two (not necessarily different) agent types $i$ and $j$, if the cooperative and defective lines $u_i^C$ and $u_j^D$ intersect at a point with the $n^C$ value in the interval $[1,n-1]$, we denote the \emph{intersection point} by $p_{ij}$ and its corresponding $n^C$ value by $n^C(p_{ij})$. 
We refer to the intersection point of an equilibrium state $x^*$, mentioned in Proposition \ref{prop11}, as the \emph{intersection point of $x^*$}.
%by $u(p_{ij})$.
If $n^C(p_{ij})$ is an integer, $p_{ij}$ is called an \emph{integer intersection point}, otherwise, a \emph{non-integer intersection point}.
%Given some $x\in\X$, if both $u^C_i$ and $u^D_j$ are active by $x$, the intersection point is said to be \emph{active (by $x$)}. %, and if they are both max at some $n^C$, it is said to be \emph{max (at $n^C$)}.
The intersection point of an equilibrium state must have an integer $n^C$-value; moreover, according to Proposition \ref{prop11}, it has to be intersected by two lines that are max active at the $n^C$ value of the state (Fig.~\ref{fig3}).
%%??
Intersection points can be classified into two types according to how the cooperative and defective lines intersect. 
An intersection point $p_{ij}$ is said to be \emph{potentially attracting} if $u_i^C(n^C(p_{ij})-1)>u_j^D(n^C(p_{ij})-1)$ since other states with $n^C$ values near $n^C(p_{ij})$ may be ``attracted to'' this type of equilibrium state; otherwise, it is said to be \emph{non-attracting} since no state with an $n^C$ value nearby will converge to this point.

\section{fluctuation sets} \label{sec:fluc}
%In this section, we need a stronger assumption on randomness, that is,
%\begin{assumption}
%	The activation sequence of agents is randomized. Every agent will activate in a finite number of time steps after its previous activation.
%\end{assumption}
As discussed in Section \ref{sec:example} and also based on the simulation results in \cite{piter2016}, the solution trajectory often does not reach an equilibrium but perpetually fluctuates. 
Then we expect the trajectory to eventually enter a set of states and never leave. %The link between these two sentences is not clear.
We define a set $\mathcal{Q}$ to be \emph{positively invariant} if for any $x(0) \in \mathcal{Q}$ and under any activation sequence, $x(t) \in \mathcal{Q}$ for all $t \geq 0$.
A positively invariant set is called minimal if it does not contain any subset that is positively invariant. 
Although seemingly trivial, for rigorousness, the following result is proved in the Appendix.
\begin{lemma}   \label{lem4}
	Given any $x(0)\in\X$ and under any random activation sequence, 
	there exists a minimal positively invariant set $\mathcal{O} \subseteq \mathcal{X}$ such that almost surely $x(T)\in\mathcal{O}$ for some $T\geq 0$.  
\end{lemma}

We note that all convergence results in this paper hold almost surely, and for simplicity, we avoid this term from the statement of the results. 
%Make sure the next sentence is clearly linked to the above lemma.
In view of Lemma \ref{lem4}, $x(t)$ reaches either an equilibrium or a non-singleton minimal positively invariant set that contains no equilibrium.
So we have the following result. 
\begin{proposition}  \label{prop:2}
	$x(t)$ never reaches an equilibrium if and only if there exists a non-singleton minimal positively invariant set $\mathcal{O} \subseteq \mathcal{X}$, such that $x(T)\in \mathcal{O}$ for some $T\geq0$.
\end{proposition}

Being in a non-singleton minimal positively invariant set, $x(t)$ will fluctuate among the states of this set and visit each one infinitely often.
The following result readily follows from the randomness of the activation sequence.
\begin{lemma}   \label{lem5} 
    Let $\mathcal{O} \subseteq \X$ be a non-singleton minimal positively invariant set.
	If $x(0) \in \mathcal{O}$, then for each $y\in\mathcal{O}$, there are infinitely many time instants $t\in\Z_{\geq0}$, at which $x(t)=y$.
\end{lemma}

Therefore, we refer to the non-singleton minimal positively invariant sets as \emph{fluctuation sets}.
%Directly from Proposition 2, we have the following proposition.
%\begin{proposition}
%	$x(t)$ always converges to an equilibrium for every initial condition $x(0)$ under imitation dynamics if and 
%  only if there is no fluctuation set. 
%\end{proposition}

Back to the graphical representation, the existence of a fluctuation set $\mathcal{O}$ implies the existence of an interval $\mathcal{I} = [\min_{x\in\mathcal{O}} n^C(x),\max_{x\in\mathcal{O}} n^C(x)]$, which we call a \emph{fluctuation interval}, such that if $x(0)\in \mathcal{O}$, then we have $n^C(x(t))\in\mathcal{I}\quad \forall t\geq 0 \text{.}$
Clearly, $\min_{x\in\mathcal{O}} n^C(x) \geq 1$ and $\max_{x\in\mathcal{O}} n^C(x)\leq n-1$ as the states with $n^C(x) = 0$ or $n$ are equilibria.
Hence, a fluctuation interval $\I$ is a sub-interval of $[1,n-1]$.
As we later show in Section \ref{Rt}, the interval $[27,63]$ in Fig.~\ref{fig1} is, indeed, a fluctuation interval. 
For trajectories that enter the corresponding fluctuation set, the number of cooperators never falls short of 27 or exceeds 63 afterwards, which implies that the max active line is cooperative every time there are 27 cooperators and defective every time there are 63 cooperators. 
This observation can be generally stated as follows. 
\begin{lemma} \label{lem6}
    If $\mathcal{I} = [a,b], a,b \in\{1,\ldots,n-1\},$ is a fluctuation interval with corresponding fluctuation set $\O$,
    then for any $x\in\mathcal{O}$ with $n^C(x) = a$ (resp. $b$), 
    among the lines active by $x$, 
    the max at $a$ (resp. $b$), is a cooperative (resp. defective) line.
\end{lemma}

Lemma \ref{lem6} states that for all $x \in \O$ with $n^C(x) = a$ (resp. $b$), the max active-by-$x$ line is cooperative (resp. defective). 
However, can the max line at $a$ be defective but never active by any $x\in\O$?
The following result provides a negative answer.
\begin{proposition} \label{prop:max}
    If $\mathcal{I} = [a,b], a,b \in\{1,\ldots,n-1\},$ is a fluctuation interval, 
    then the max line at $a$ (resp. $b$) is a cooperative (resp. defective) line.
\end{proposition}

The graphical representation of the imitation dynamics for Fig.~\ref{fig1}, is shown in Fig.~\ref{fig7}.
As seen, the max lines at $n^C = 27$ and $63$ are $u^C_4$ and $u^D_3$, which are cooperative and defective, respectively.

\begin{figure}
	\centering
	\begin{tikzpicture}
		\begin{axis}[
			standard,
			xlabel=$n^C$,
			xtick={15.1,51.8,38.8,66},
			ylabel=$u$,
			legend pos=outer north east,
			legend style={font=\small},
			every axis y label/.style={
    		at={(ticklabel* cs:1.04)},
    		anchor= east,
}
		]
		\addplot [
			name path=P1,
			domain=0:100, 
			samples=10, 
			color=blue,
			]coordinates
			{(0,3.55)
			(100,2)};
		\addlegendentry{$u^C_i$}
		
		\addplot [
			name path=P2,
			domain=0:100, 
			samples=10, 
			color=blue,
			style=dashed
			]coordinates
			{(0,2.5)
			(100,3)};
		\addlegendentry{$u^D_i$}

		]
		\addplot [
			name path=P3,
			domain=0:100, 
			samples=10, 
			color=red,
			]coordinates
			{(0,3)
			(100,0)};
		\addlegendentry{$u^C_j$}
		
		\addplot [
			name path=P4,
			domain=0:100, 
			samples=10, 
			color=red,
			style=dashed
			]coordinates
			{(0,0.75)
			(100,3.5)};
		\addlegendentry{$u^D_j$}

		\path [name intersections={of=P3 and P2,by=A}];
		\node [circle,fill=black,inner sep=1.5pt,label=90:$p_{ji}$] at (A) {};
		\path [name intersections={of=P1 and P2,by=B}];
		\node [circle,fill=black,inner sep=1.5pt,label=90:$p_{ii}$] at (B) {};
		\path [name intersections={of=P3 and P4,by=C}];
		\node [circle,fill=black,inner sep=1.5pt,label=90:$p_{jj}$] at (C) {};
		\path [name intersections={of=P1 and P4,by=D}];
		\node [circle,fill=black,inner sep=1.5pt,label=-90:$p_{ij}$] at (D) {};
		%Here the blue parabloa is defined
		%\draw [black, fill] (16.5,82) circle (2pt) node [right] {$a$};
		%\fill[black,name intersections={of=P1 and P2,total=\t}] ;
		%\path [name intersections={of=P1 and P2,by={CS}}];
		%	\foreach \s in {1,...,\t}{(intersection-\s) circle (2pt) node {\footnotesize\s}};
		%\node[red,scale=3] at (intersection of  line 1 and line 2){.};
		
		%\draw [black, fill] (50.5 ,148) circle (2pt) node [right] {$b$};
		%\draw [black, fill] (70 ,190) circle (2pt) node [right] {$c$};
		%\node[dot=E] (E) at (CS) {};

		\draw [gray, dashed] (A) -- (15.1,0);
		\draw [gray, dashed] (B) -- (51.8,0);
		\draw [gray, dashed] (C) -- (38.8,0);
		\draw [gray, dashed] (D) -- (66,0);

		\end{axis}
	\end{tikzpicture}
	\caption{\textbf{Fluctuation point.}
	The point $p_{ii}$ is a fluctuation point since it is a non-integer potentially attracting intersection point, there exists some state $y\in\X$ such that among the lines active by $y$, $u^C_i$ is max at $n^C(y)=\floor{n^C(p_{ii})}=51$, and there exists some state $z\in\X$ such that among the lines active by $z$, $u^D_i$ is max at $n^C(z) = \ceil{n^C(p_{ij})}=52$.
	The point $p_{ji}$ is a fluctuation point if $u^C_i$ can be inactive when $n^C(t) = \floor{n^C(p_{ji})}$ and when $n^C(t+1) = \ceil{n^C(p_{ji})}$.
	The point $p_{jj}$ can never be a fluctuation point since the max active line can never be $u^C_j$ and $u^D_j$ at $n^C=\floor{n^C(p_{jj})}=38$ and $n^C = \ceil{n^C(p_{jj})}=39$, respectively. 
	An example of a fluctuation around $n^C(p_{ii})$ is described as follows: if at some time $t$ we have $n^C(t)=\floor{n^C(p_{ii})}=51$, both $u^C_i$ and $u^D_i$ are active, and the active agent is a defector, then she will switch to cooperation and by moving one unit right on the $n^C$-axis, we obtain $n^C(t+1)=\ceil{n^C(p_{ii})}=52$. 
	If the active agent at $t+1$ is a cooperator, then by moving one unit left on the $n^C$-axis, we obtain $n^C(t+2)=\floor{n^C(p_{ii})}=51$. 
	The fluctuation continues until one of $u^C_i$ and $u^D_i$ becomes inactive.} 
	\label{fig4}
\end{figure}
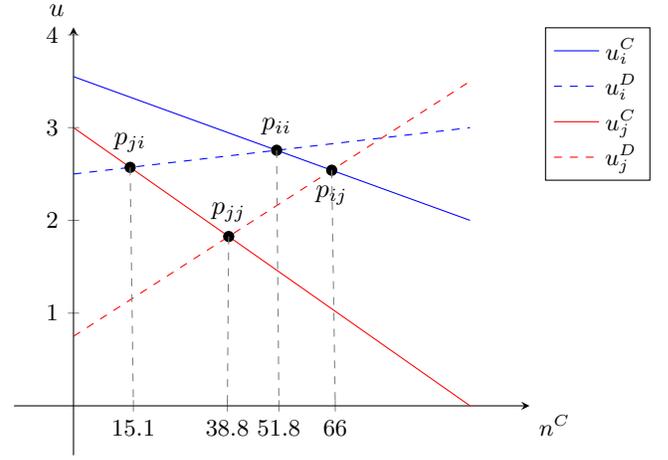

Interestingly, when $n^C$ is in any of the intervals $[27,28]$ and $[62,63]$, it fluctuates back and forth several times and then moves towards the center of the fluctuation interval. 
This, intuitively, implies the existence of two intersection points with $n^C$-values in the interval $[27,28]$ and $[62,63]$ that cause the fluctuation.
Such an intersection point, say $p_{ij}$, has the property that for some time $t$ when $n^C(t)=\floor{n^C(p_{ij})}$ and a defector is active, $u^C_i$ is the maximum among all active lines at $\floor{n^C(p_{ij})}$, resulting in $n^C(t+1) = \ceil{n^C(p_{ij})}$, where $\floor{x}$ (resp. $\ceil{x}$) is the greatest (resp. smallest) integer less than (resp. greater than) or equal to real number $x$.
So the intersection point is potentially attracting.
Moreover, at $t+1$ a defector is active and $u^D_j$ is the maximum among all active lines at $\ceil{n^C(p_{ij})}$, resulting in  $n^C(t+2)=n^C(t)=\floor{n^C(p_{ij})}$. 
This repetitive behavior of $n^C$ can last until either $u^C_i$ or $u^D_j$ become inactive or they no longer excel the active lines. 
This motivates us to make the following definition.
Given a state $x\in\X$, let $\X_{+1}(x)$ denote the set of states whose entries are all the same as those of $x$, except for a single entry that is one more:
\begin{align*}
    \X_{+1}(x) =
        \big\{ z\in\X\,|\, &\exists k\in\{1,\ldots, m\}:   \\
             &(z_i = x_i \forall i\neq k, z_k = x_k +1)\big\}.
\end{align*} 
We call a non-integer potentially attracting intersection point $p_{ij}$ of type-$i$ and type-$j$ agents a \emph{fluctuation point}, if there exist both a state $x\in\X$ with $n^C(x)=\floor{n^C(p_{ij})}$ such that among the lines active by $x$, $u^C_i$ is max at $\floor{n^C(p_{ij})}$,
and a state $y\in\X_{+1}(x)$ with $n^C(y) = \ceil{n^C(p_{ij})}$ such that among the lines active by $y$, $u^D_j$ is max at $\ceil{n^C(p_{ij})}$ (Fig.~\ref{fig4}). 
We say an interval `contains' an intersection point if the $n^C$-value of that intersection point belongs to the interval.
%Therefore, we can formulated the following lemma.
%%

\begin{lemma} \label{fluct_pt}
    If $\mathcal{I} = [a,b],a,b \in\{1,\ldots,n-1\},$ is a fluctuation interval, then each of the intervals $(a,a+1)$ and $(b-1,b)$ contains a fluctuation point.
\end{lemma}

Back to the example, the non-integer potentially attracting intersection points $p_{33}$ and $p_{34}$ are fluctuation points with $n^C(p_{33}) \in (27,28)$ and $n^C(p_{34}) \in(62,63)$ (Fig.~\ref{fig7}).

We know that the $n^C$-value of a trajectory starting from the fluctuation set can never ``pass'' the end points of the fluctuation interval.
Namely, given a fluctuation set $\O$ with the fluctuation interval $\I = [a,b]$, if $x(0)\in \O$, then $n^C(x(t)) \in[a,b]\ \forall t\geq0$.
Apparently, the two end intervals $[a,a+1]$ and $[b-1,b]$ perform as a ``wall'' that do not allow the $n^C$-value of any solution trajectory to pass them from right and left, respectively.
Formally, we define an interval $[a,a+1], a \in \{1,\ldots,n-1\},$ that contains a fluctuation point to be a \emph{left wall} if for any trajectory $x(t)$ with $n^C(x(0)) \geq a+1$, we have $n^C(x(t)) \geq a\ \forall t\geq0$ under any activation sequence.
Similarly, we define a \emph{right wall}.
Now we reveal the necessary and sufficient condition for an interval to be a wall in terms of the graphical representation. 
For the interval $[a,a+1]$, denote the max defective lines at $a$ and $a+1$ by $u^D_r$ and $u^D_s$, respectively. 
Then let $\mathcal{R}\subseteq\{1,\ldots,n\}$ denote the set of agents whose cooperative lines are below the defective line $u^D_r$ at $a$, and let $\mathcal{S}\subseteq\{1,\ldots,n\}$ denote the set of agents whose cooperative lines are above the defective line $u^D_s$ at $a+1$ (Fig.~\ref{leftwall}).
\begin{lemma} \label{wall}
    An interval $[a,a+1], a \in \{1,\ldots,n-2\},$ containing a fluctuation point is a left wall if and only if $|\mathcal{R}\setminus \mathcal{S}| < a$.
\end{lemma}

A similar result to that of Lemma \ref{wall} can be stated for a right wall. 
For the interval $[b-1,b], b\in\{2,\ldots,n-1\}$, denote the max cooperative lines at $b-1$ and $b$ by $u^C_l$ and $u^C_k$, respectively. 
Then let $\mathcal{L}\subseteq\{1,\ldots,n\}$ denote the set of agents whose defective lines are above the cooperative line $u^C_l$ at $b-1$, and let $\mathcal{K}\subseteq\{1,\ldots,n\}$ denote the set of agents whose defective lines are below the cooperative line $u^C_k$ at $b$.
\begin{lemma} \label{lem:rightWall}
    An interval $[b-1,b], b \in \{2,\ldots,n-1\},$ containing a fluctuation point is a right wall if and only if $|\mathcal{K}\setminus \mathcal{L}| < n-b$.
\end{lemma}

Using Lemma \ref{wall} and \ref{lem:rightWall}, we prove the following theorem states that each fluctuation interval consists of a left and right wall.
\begin{theorem} \label{endinter}
    If $\mathcal{I} = [a, b],a,b \in \{1,\ldots,n-1\},$ is a fluctuation interval, then $[a,a+1]$ is a left wall and $[b-1,b]$ is a right wall.
\end{theorem}

According to Theorem \ref{endinter}, the total number of cooperators is confined to a fluctuation interval $\I = [a,b]$ not only for initial states within the corresponding fluctuation set, i.e., $x(0)\in\O$, but also for any other initial state with the number of cooperators in the open fluctuation interval, i.e., $x(0)\not\in\O, n^C(x(0))\in[a+1,b-1]$.
\begin{corollary} \label{constraint}
    If $\I = [a,b]$, where $a,b \in \{1, \ldots, n-1\}$, is a fluctuation interval, then for any initial condition $x(0)$ with $n^C(x(0)) \in [a+1,b-1]$, we have $n^C(x(t)) \in \I ~\forall t>0$.
\end{corollary}

This result also puts forward the set 
\begin{equation} \label{eq:Q}
    \mathcal{Q} = \left\{x\in\X\ | \, n^C(x) \in [a+1,b-1]\right\}
\end{equation}
as a subset of the \emph{basin of attraction} of the union of all fluctuation sets with the corresponding fluctuation interval $\I = [a,b]$.
Moreover, we now have a sufficient condition for the solution trajectory to equilibrate, thanks to Proposition \ref{prop:2}.
\begin{corollary} \label{c2}
    $x(t)$ reaches an equilibrium if there is no interval $\I = [a,b],a,b\in\{1,\ldots,n-1\}$ such that $[a,a+1]$ is a left and $[b-1,b]$ is a right wall.
\end{corollary}
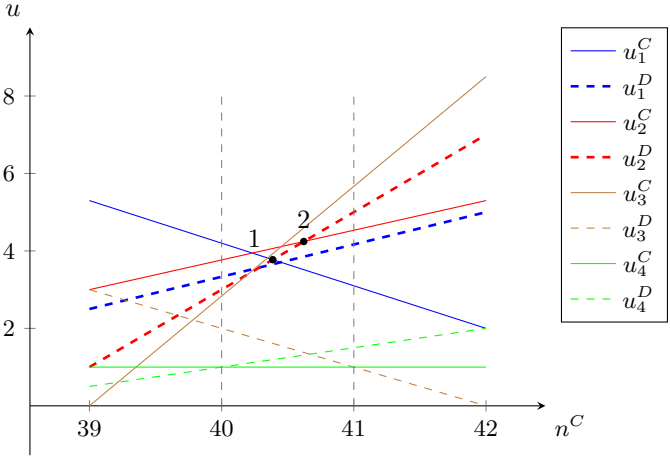
\begin{figure} 
	\centering
	\begin{tikzpicture} 
		\begin{axis}[
			standard,
			xlabel=$n^C$,
			xtick={39,40,41,42},
			ylabel=$u$,
			legend pos=outer north east,
			legend style={font=\small},
			every axis y label/.style={
    		at={(ticklabel* cs:1.04)},
    		anchor= east,
}
		]
		\addplot [
			name path=P1,
			domain=0:100, 
			samples=10, 
			color=blue,
			]coordinates
			{(39,5.3)
			(42,2)};
		\addlegendentry{$u^C_1$}
		
		\addplot [
			name path=P2,
			domain=0:100, 
			samples=10, 
			color=blue,
			line width=1pt,
			style=dashed
			]coordinates
			{(39,2.5)
			(42,5)};
		\addlegendentry{$u^D_1$}

		\addplot [
			name path=P3,
			domain=0:100, 
			samples=10, 
			color=red,
			]coordinates
			{(39,3)
			(42,5.3)};
		\addlegendentry{$u^C_2$}
		
		\addplot [
			name path=P4,
			domain=0:100, 
			samples=10, 
			color=red,
			line width=1pt,
			style=dashed
			]coordinates
			{(39,1)
			(42,7)};
		\addlegendentry{$u^D_2$}
		
		\addplot [
			name path=P5,
			domain=0:100, 
			samples=10, 
			color=brown,
			]coordinates
			{(39,0)
			(42,8.5)};
		\addlegendentry{$u^C_3$}
		
		\addplot [
			name path=P6,
			domain=0:100, 
			samples=10, 
			color=brown,
			style=dashed
			]coordinates
			{(39,3)
			(42,0)};
		\addlegendentry{$u^D_3$}
		
		\addplot [
			name path=P7,
			domain=0:100, 
			samples=10, 
			color=green,
			]coordinates
			{(39,1)
			(42,1)};
		\addlegendentry{$u^C_4$}
		
		\addplot [
			name path=P8,
			domain=0:100, 
			samples=10, 
			color=green,
			style=dashed
			]coordinates
			{(39,0.5)
			(42,2)};
		\addlegendentry{$u^D_4$}

		\path [name intersections={of=P1 and P4,by=A}];
 		\node [circle,fill=black,inner sep=1pt,label=125:$1$] at (A) {};
% 		\path [name intersections={of=P1 and P2,by=B}];
% 		\node [circle,fill=black,inner sep=1pt,label=270:$2$] at (B) {};
 		\path [name intersections={of=P3 and P4,by=C}];
 		\node [circle,fill=black,inner sep=1pt,label=90:$2$] at (C) {};
% 		\path [name intersections={of=P1 and P4,by=D}];
% 		\node [circle,fill=black,inner sep=1.5pt,label=-90:$p_{ij}$] at (D) {};
		%Here the blue parabloa is defined
		%\draw [black, fill] (16.5,82) circle (2pt) node [right] {$a$};
		%\fill[black,name intersections={of=P1 and P2,total=\t}] ;
		%\path [name intersections={of=P1 and P2,by={CS}}];
		%	\foreach \s in {1,...,\t}{(intersection-\s) circle (2pt) node {\footnotesize\s}};
		%\node[red,scale=3] at (intersection of  line 1 and line 2){.};
		
		%\draw [black, fill] (50.5 ,148) circle (2pt) node [right] {$b$};
		%\draw [black, fill] (70 ,190) circle (2pt) node [right] {$c$};
		%\node[dot=E] (E) at (CS) {};
% 
% 		\draw [gray, dashed] (A) -- (15.1,0);
% 		\draw [gray, dashed] (B) -- (51.8,0);
 		\draw [gray, dashed] (40,0) -- (40,8);
		\draw [gray, dashed] (41,0) -- (41,8);

		\end{axis}
	\end{tikzpicture}
	\caption{\textbf{A left wall.} 
	The interval $[40,41]$ is a left wall containing two fluctuation points labelled by $1$ and $2$.
	The type distribution vector $p$ of this population is $(30,30,20,20)$.
	The max defective line is $u_1^D$ at $n^C = 40$ and $u_2^D$ at $n^C = 41$.
	The set of agents whose types have cooperative lines below $u_1^D$ at $n^C = 40$ is $\mathcal{R} = \{61,\ldots,100\}$, which consists of \textcolor{brown}{type-$3$} and \textcolor{green}{type-$4$} agents and the set of agents whose types have cooperative lines above $u_2^D$ at $n^C = 41$ is $\mathcal{S} = \{61,\ldots,80\}$, which only consists of \textcolor{brown}{type-$3$} agents. 
	Hence, $|\mathcal{R}\setminus \mathcal{S}| < 40$, which in view of Lemma \ref{wall}, implies that for any trajectory $x(t)$ with $n^C(0) \geq 41$, we have $n^C(x(t)) \geq 40 \;\forall t\geq0$.}
	\label{leftwall}
\end{figure}

However, having both a left and a right wall does not guarantee the existence of a fluctuation interval, and hence, set.
First of all, there may by another left or right wall within the interval defined by the two walls, making the corresponding set not minimal. 
Secondly, even if the interval defined by the two walls does not include any other wall, it may contain an equilibrium point, which may, again, make the corresponding set not minimal.
By ignoring the second case, we can provide the following result.
\begin{proposition} \label{th2}
    An interval $\I = [a,b],a,b\in\{1,\ldots,n-1\}$ is a fluctuation interval if it is the minimal interval with the properties that $[a,a+1]$ is a left wall, $[b-1,b]$ is a right wall, and for all equilibrium states $x^*\in\X^*$, we have $n^C(x^*) \notin \mathcal{I}$.
\end{proposition}

By using Lemmas \ref{wall} and \ref{lem:rightWall}, we acquire the following sufficient condition on the graphical representation for having a fluctuation interval. 
\begin{theorem} \label{th2_pts}
    An interval $\I\subseteq[1,n-1]$ is a fluctuation interval if there are two fluctuation points $p_{ij}$ and $p_{lk}$ of agent types $i$, $j$, $l$, and $k$, such that the followings are true:
    \begin{enumerate}
        \item $\mathcal{I} = \left[\floor{n^C(p_{ij})},\ceil{n^C(p_{lk})}\right]$;
	    \item for interval $[\floor{n^C(p_{ij})},\floor{n^C(p_{ij})}+1]$, we have $|\mathcal{R}\setminus \mathcal{S}| < \floor{n^C(p_{ij})}$, and for interval $[\ceil{n^C(p_{lk})}-1,\ceil{n^C(p_{lk})}]$, we have $|\mathcal{K}\setminus \mathcal{L}| < n-\ceil{n^C(p_{lk})}$.
	    \item no subinterval of $\mathcal{I}$ satisfies the above conditions;
		\item for all equilibrium states $x^*$, we have $n^C(x^*) \notin \mathcal{I}$.
    \end{enumerate}
\end{theorem}

The above conditions ensure that an interval $\I$ is a fluctuation interval, and hence, corresponding to which, there is at least one fluctuation set $\O$.
Although the theorem does not explicitly determine the form of $\O$, it specifies the possible range of its number of cooperators, i.e., $\I$.

\section{Stability Analysis} \label{sec:stability}
We analyze the stability of the population dynamics under the imitation update rule $\eqref{updateRule}$.
Let $\|\cdot\|$ denote the one-norm, i.e., $\|x\| = \sum_{i=1} |x_i|$.
An equilibrium state $s\in\X^*$ is \emph{(Lyapunov) stable} if for every open ball (in the state space) $\mathcal{B}_\epsilon(s) = \{x\in\X\,|\, \|x-s\|<\epsilon\}$, there exists a ball $\mathcal{B}_\delta(s)$ such that for every $x(0)\in\mathcal{B}_\delta$, we have $x(t)\in \mathcal{B}_\epsilon\ \forall t \geq 0$ under any activation sequence. 
Due to the discrete nature of the state space $\mathcal{X}$, we require $\epsilon$ and $\delta$ to be at least 2.
Otherwise, if we take $\delta = 1$, we have $\mathcal{B}_\delta(s) = \{s\}$, resulting in $x(t) = s\ \forall t \geq 0$, even if $s$ is unstable. 
Also, if we take $\epsilon = 1$, we have $\mathcal{B}_\epsilon(s) = \{s\}$, requiring $x(t) = s\ \forall t\geq0$ for $s$ to be stable--an impossible condition.
An equilibrium state $s\in\X^*$ is \emph{asymptotically stable} if it is stable and there exists some ball $\mathcal{B}_\delta(s)$ such that $\text{lim}_{t\rightarrow{\infty}}||x(t)-s||=0$ for all $x(0) \in \mathcal{B}_\delta(s)$ and under any random activation sequence.

First, we study populations with only one type of agents, i.e., $m=1$.
\begin{proposition} \label{p5}
    In a population with only one type of agents, an equilibrium state $s=(x_1), x_1\in\{1,\ldots,n-1\}$, is stable if and only if $u^C_1(x_1 - 1) \geq u^D_1(x_1 - 1)$ and $u^C_1(x_1+1) \leq u^D_1(x_1+1)$, and is asymptotically stable if and only if the inequalities are strict, i.e., $u^C_1(x_1 - 1) > u^D_1(x_1 - 1)$ and $u^C_1(x_1+1) < u^D_1(x_1+1)$.
\end{proposition}

Now we proceed to populations with at least two types. 
We start with the extreme equilibria and then proceed to other equilibria.
\begin{proposition} \label{p6}
    In a population with at least two types of agents, 
    the equilibrium state $(0,\ldots,0)$ (resp. $(n_1,\ldots,n_m)$) is unstable if and only if there exists an agent type $i$ such that if $n_i > 1$ (resp. $n_i<n$), then 
    \begin{equation*}
        u_i^C(1)>u_j^D(1) \text{ (resp. } u_i^D(n-1)>u_j^C(n-1)\text{)} \quad \forall j,
    \end{equation*} 
    and if $n_i = 1$ (resp. $n_i = n$), then
    \begin{equation*}
        u_i^C(1)>u_j^D(1) \text{ (resp. } u_i^D(n-1)>u_j^C(n-1)\text{)} \quad \forall j\neq i.
    \end{equation*} 
\end{proposition}

If we exclude populations having a type with just one agent, we obtain the following result

\begin{theorem} \label{st1}
    In a population with at least two types of agents, and where every type has at least two agents, 
    the extreme equilibrium state $(0,\ldots,0)$ (resp. $(n_1,\ldots,n_m)$) is stable if and only if
    among the max lines at $1$ (resp. $n-1$), at least one is defective (resp. cooperative), 
    and is asymptotically stable if and only if 
    the max line at $1$ (resp. $n-1$) is defective (resp. cooperative).
\end{theorem}

%%%
\begin{theorem} \label{st2}
    In a population with at least two types of agents, a non-extreme equilibrium state $x^*$ is stable if and only if every state $y\in
    \{x\in\X~|~||x^*-x||=1\}$ is an equilibrium.
    Moreover, non-extreme equilibria are not asymptotically stable.
\end{theorem}

    According to the theorem, the only possibility for a non-extreme equilibrium to be stable is to be fully surrounded by other equilibria.
    Namely, if an equilibrium $x^*$ with $n^C(x^*) = a$ is stable, then all of the states at $n^C = a-1$ and $n^C = a+1$ must also be equilibria.
    In view of the graphical representation, a population is unlikely to possess such a feature.
    Moreover, possessing the two extreme equilibria, the imitation dynamics never admit a globally asymptotically stable state as this requires the state to be the only equilibrium. 
    %Therefore, for a general population, one may roughly confine the possible stable equilibrium states to the two extremes. 
    %This implies that if the update rule is noisy, meaning that the agents sometimes randomly pick a strategy, \tB{then the solution trajectory eventually reaches an invariant set}.  

%%%%%%%%%%%%%%%%%%%%%%%%%%%%%%%%%%%%%%%%%%%%%%%
%
\begin{figure*}[t]
	\centering
	\begin{tikzpicture}
		\begin{axis}[
			standard,
			width=12cm,
			height=7cm,
			legend pos=outer north east,
			legend style={font=\small},
			xlabel=$n^C$,
			xtick={0,15,27.5,37.5,62.5,69.4,75,79.6,88.5,90,100},
			extra x ticks={90.9},
			extra x tick labels={$90.9$},
			extra x tick style={% changes for extra x ticks
        tick label style={xshift=1mm}
    },
% 			xticklabel style = {font=\tiny}
            %label style={font=\tiny},
            %tick label style={font=\tiny},
            xticklabel style={rotate=270,font=\tiny},
			ylabel=$u$,
    		axis lines=middle, 
    		domain=-5:100,
    		samples=100,
    		xmin=0, xmax=100,
			ymin=0, ymax=120,
			every axis y label/.style={
    		at={(ticklabel* cs:1.1)},
    		anchor=east},
    		every axis x label/.style={
    		at={(ticklabel* cs:1.1)},
    		anchor=east
}
		]
		\addplot [
			name path=P1,
			domain=0:100, 
			samples=10, 
			color=red,
			] 
			{-0.67*x+36.85};
		\addlegendentry{$u_1^C$}

		\addplot [
			name path=P2,
			domain=0:100, 
			samples=10, 
			color=red,
			style=dashed
			] 
			{1.5*x-30};
		\addlegendentry{$u_1^D$}
		%Here the blue parabloa is defined

		\addplot [
			name path=P3,
			domain=0:100, 
			samples=10, 
			color=green,
			] 
			{6.8*x-512};
		\addlegendentry{$u_2^C$}

		\addplot [
			name path=P4,
			domain=0:100, 
			samples=10, 
			color=green,
			style=dashed
			] 
			{2*x-130};
		\addlegendentry{$u_2^D$}

		\addplot [
			name path=P5,
			domain=0:100, 
			samples=10, 
			color=brown,
			] 
			{-0.4*x+50};
		\addlegendentry{$u_3^C$}

		\addplot [
			name path=P6,
			domain=0:100, 
			samples=10, 
			color=brown,
			style=dashed
			] 
			{0.976*x+12.16};
		\addlegendentry{$u_3^D$}

		\addplot [
			name path=P7,
			domain=0:100, 
			samples=10, 
			color=blue,
			] 
			{-1.25*x+95.635};
		\addlegendentry{$u_4^C$}

		\addplot [
			name path=P8,
			domain=0:100, 
			samples=10, 
			color=blue,
			style=dashed
			] 
			{2.495*x-130.9375};
		\addlegendentry{$u_4^D$}
		%Here the blue parabloa is defined

		\path [name intersections={of=P1 and P2,by=A}];
		\node [circle,fill=black,inner sep=1.5pt,label=90:$p_{11}$] at (A) {};
		\path [name intersections={of=P1 and P6,by=B}];
		\node [circle,fill=black,inner sep=1.5pt,label=90:$p_{13}$] at (B) {};
		\path [name intersections={of=P5 and P6,by=C}];
		\node [circle,fill=black,inner sep=1.5pt,label=90:$p_{33}$] at (C) {};
		\path [name intersections={of=P7 and P6,by=D}];
		\node [circle,fill=black,inner sep=1.5pt,label=90:$p_{43}$] at (D) {};
		\path [name intersections={of=P2 and P7,by=E}];
		\node [circle,fill=black,inner sep=1.5pt,label=90:$p_{41}$] at (E) {};
		\path [name intersections={of=P2 and P5,by=F}];
		\node [circle,fill=black,inner sep=1.5pt,label=270:$p_{31}$] at (F) {};
		\path [name intersections={of=P3 and P6,by=G}];
		\node [circle,fill=black,inner sep=1.5pt,label=180:$p_{23}$] at (G) {};
		\path [name intersections={of=P5 and P8,by=H}];
		\node [circle,fill=black,inner sep=1.5pt,label=120:$p_{34}$] at (H) {};
		\path [name intersections={of=P7 and P8,by=I}];
		\node [circle,fill=black,inner sep=1.5pt,label=180:$p_{44}$] at (I) {};
		\path [name intersections={of=P1 and P8,by=J}];
		\node [circle,fill=black,inner sep=1.5pt,label=90:$p_{14}$] at (J) {};
		\path [name intersections={of=P7 and P4,by=K}];
		\node [circle,fill=black,inner sep=1.5pt,label=180:$p_{42}$] at (K) {};
		\path [name intersections={of=P3 and P4,by=L}];
		\node [circle,fill=black,inner sep=1.5pt,label=100:$p_{22}$] at (L) {};
		\path [name intersections={of=P5 and P4,by=M}];
		\node [circle,fill=black,inner sep=1.5pt,label=90:$p_{32}$] at (M) {};
		\path [name intersections={of=P3 and P8,by=N}];
		\node [circle,fill=black,inner sep=1.5pt,label=180:$p_{24}$] at (N) {};
		\path [name intersections={of=P2 and P3,by=O}];
		\node [circle,fill=black,inner sep=1.5pt,label=90:$p_{21}$] at (O) {};

		\draw [gray, dashed] (B) -- (15,0);
		\draw [gray, dashed] (M) -- (75,0);
		\draw [gray, dashed] (G) -- (90,0);

		\draw [gray, dashed] (C) -- (27.5,0);
		\draw [gray, dashed] (D) -- (37.5,0);
		\draw [gray, dashed] (H) -- (62.5,0);
		\draw [gray, dashed] (K) -- (69.4,0);
		\draw [gray, dashed] (L) -- (79.6,0);
		\draw [gray, dashed] (N) -- (88.5,0);
		\draw [gray, dashed] (O) -- (90.9,0);

		\draw [draw=none, fill=yellow, fill opacity=0.1] (27,0) rectangle (63,120);
		\draw [draw=none, fill=green, fill opacity=0.1] (37,0) rectangle (63,110);

 		\end{axis}
	\end{tikzpicture}
	%\captionsetup{font=tiny}
	\caption{The graphical representation of a well-mixed population of $100$ agents, in which there are $4$ agent types. 
	The type distribution is $(13,18,5,64)$. 
	The integer intersection points are $p_{13}$, $p_{32}$, $p_{23}$, and the non-integer but potentially attracting intersection points are $p_{33}$, $p_{11}$, $p_{43}$, $p_{31}$, $p_{41}$, $p_{14}$, $p_{44}$, $p_{34}$, $p_{42}$.
	Among them, $p_{13}$ and $p_{23}$ are equilibrium points, and $p_{33}$, $p_{43}$, $p_{31}$, $p_{41}$, and $p_{34}$ are fluctuation points.
	We have left walls $[\floor{n^C(p_{33})}, \ceil{n^C(p_{33})}]$ and $[\floor{n^C(p_{43})}, \ceil{n^C(p_{43})}]$, and right wall $[\floor{n^C(p_{34})}, \ceil{n^C(p_{34})}]$. 
	However, due to minimality, the only fluctuation interval is $[\floor{n^C(p_{43})}, \ceil{n^C(p_{34})}] = [37,63]$, which is colored by light green.} 
	\label{fig7}
\end{figure*}
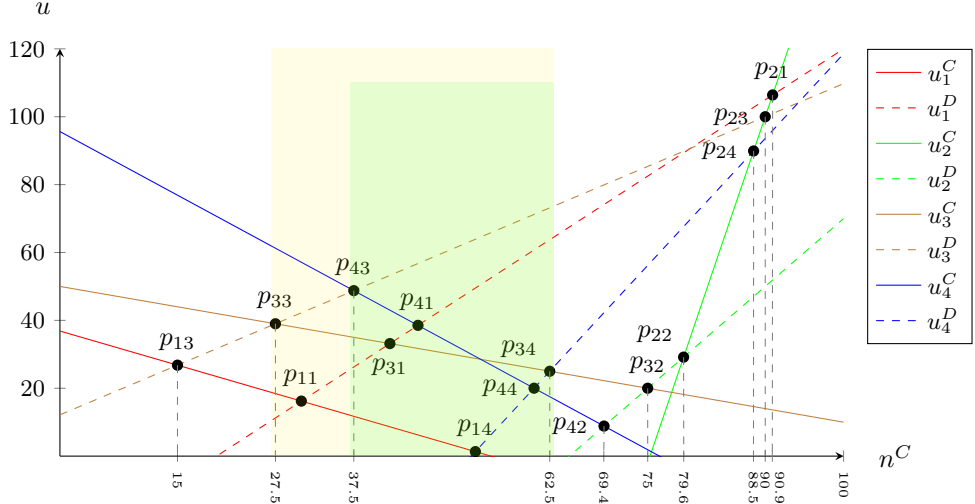

\section{Revisiting the example} \label{Rt}
First, we find the cooperative and defective lines of each type of agent:
\begin{equation*}		
	\begin{cases}
		u_1^C = 	-0.7n^C + 36.9  \\
		u_1^D = 	1.5n^C -	30 	
	\end{cases}
	\begin{cases}
		u_2^C = 	6.8n^C -512 \\
		u_2^D = 	2n^C -130 
	\end{cases}
\end{equation*}
\begin{equation*}		
	\begin{cases}
		u_3^C = 	-0.4n^C + 50  \\
		u_3^D = 	n^C + 12.2 		
	\end{cases}
	\begin{cases}
	    u_4^C = 	-1.3n^C+ 95.7  \\
		u_4^D = 	2.5n^C-	130.9 	
	\end{cases}.
\end{equation*}
Next, we find the intersection points between all pairs of cooperative and defective lines in order to determine the equilibria and fluctuation sets. 
Among all intersection points, we have integer intersection points $p_{13}$, $p_{32}$, $p_{23}$, which are candidates for equilibrium states, and non-integer but potentially attracting intersection points $p_{33}$, $p_{11}$, $p_{43}$, $p_{31}$, $p_{41}$, $p_{14}$, $p_{44}$, $p_{34}$, $p_{42}$, which are candidates for fluctuation points. 

For the equilibria, the states corresponding to $p_{32}$ cannot be an equilibrium according to \eqref{property} since $n^C(p_{32}) = 75$ and $n_2=18<25 = n-75 = n^D$, which implies that at least one of the defective lines above $p_{32}$ is active at $n^C = 75$.
In view of Theorem \ref{th1}, the equilibrium states corresponding to $p_{13}$ and $p_{23}$ are $(x_1,x_2,0,0)$ where $x_1,x_2\in\mathbb{N}$, $x_1>0$, $x_1+x_2=15$, and $(13,y_2,y_3,y_4)$ where $y_2,y_3,y_4\in\mathbb{N}$, $y_2>0$, $y_3<5$, $y_2+y_3+y_4 = 77$. 
These states together with the extreme equilibria $(0,0,0,0)$ and $(13,18,5,64)$ form all of the equilibrium states of the dynamics. 
Namely, $\X^*$ equals
\begin{align*}
        \big\{ (0,0,0,0),   (13,18,5,64), (x_1,x_2,0,0), (13,y_2,y_3,y_4) |&  \\
               x_1,x_2,y_2,y_3,y_4 \in \mathbb{N}, x_1\geq1, x_1+x_2=15, & \\
               y_2\geq1,y_3\leq 4, y_2+y_3+y_4 = 77& \big\}.
\end{align*}

As with the fluctuation points, the intersection point $p_{11}$ cannot be one as both of the cooperative and defective lines $u^C_3$ and $u^D_3$ are above $p_{11}$ at $n^C(p_{11})$.
The same holds for $p_{14}$, $p_{44}$ and $p_{42}$.
So the only possible fluctuation points among the candidates are $p_{33}$, $p_{43}$, $p_{31}$, $p_{41}$, and $p_{34}$. 
By checking the conditions in Lemmas \ref{wall} and \ref{lem:rightWall}, we find that intervals $[\floor{p_{33}},\ceil{p_{33}}]$ and $[\floor{p_{43}}, \ceil{p_{43}}]$ are left walls and interval $[\floor{p_{34}},\ceil{p_{34}}]$ is a right wall.
So due to minimality, the only fluctuation interval is $[\floor{p_{43}},\ceil{p_{34}}] = [37, 63]$, which can be also verified by Theorem \ref{th2_pts}.
Therefore, there exist one (or more) fluctuation set(s) $\O$ with $\min_{x\in\mathcal{O}} n^C(x) = 37$ and $\max_{x\in\mathcal{O}}n^C(x) = 63$.

Now, thanks to Corollary \ref{constraint}, a trajectory $x(t)$ never converges to an equilibrium if $n^C(x(T))\in[38,62]$ at some time $T$. 
This explains the behavior of trajectories $x^1(t)$, $x^2(t)$, $x^3(t)$ in Fig.~\ref{fig1}. 
Since the $n^C$-values of these four trajectories enter the fluctuation interval, they never converge to the equilibrium at $n^C= 15$ like $x^4(t)$ does. 
Note that the $n^C$ values of trajectories trapped in the fluctuation set(s) seem to be constrained in the interval $[37,38]$ is because the probability of leaving the fluctuation at this interval is fairly small, although greater than zero. 
Similarly, the probability of leaving the fluctuation at $[45, 46]$ from the right end is much smaller than leaving from the left end, so we rarely see $n^C$ reaching 63, even though it does.
More generally, we have the following cases regarding $n^C(0)$:
\begin{case}
	For $0<n^C(0)<27$, if $x_1(0)+x_2(0) \geq 1$, then the max active line is always a cooperative line at $n^C(0)$. 
	Thus, $n^C(x(t))$ will increase until it enters the fluctuation interval at some time $T$, resulting in $x(t) \in  \mathcal{O} \forall t\geq T$.
	If $x_1(0) = x_2(0) = 0$, then depending on the activation sequence, $x(t)$ may either reach an equilibrium corresponding to $p_{13}$ or enter the fluctuation set.
\end{case}
\begin{case}
	For $27\leq n^C(0) \leq 37$, if $x_4(0) \geq 1$, then $x(t) \in  \mathcal{O} \forall t\geq T$ for some time $T\geq0$.
	If $x_4(0) = 0$, then the $n^C(x(t))$ fluctuates between $27$ and $28$ until a type-$4$ defector switches to cooperation, making $u^C_4$ active, driving the solution trajectory to the fluctuation set.
	Namely, again $x(t) \in  \mathcal{O} \forall t\geq T$ for some time $T\geq0$.
\end{case}
\begin{case}
	For $38\leq n^C(0) \leq 62$, $x(t) \in  \mathcal{O}\forall t\geq0$.
\end{case}
\begin{case}
	For $63\leq n^C(0) \leq 81$, the max active line is always a defective line at $n^C(0)$; otherwise, all agents of types $1,3$ and $4$ must be cooperative, resulting in $n^C(0) \geq n_1+n_3+n_4 = 82 > 81$.
	Therefore, for some time $T$, 
	$x(t) \in  \mathcal{O}$  $\forall t\geq T$.
\end{case}
\begin{case}
	For $82\leq n^C(0) \leq 89$, if $x_1(0)\leq 12$ or $x_3(0) \leq 4$ or $x_4(0) \leq 63$, then the
	max active line is a defective line at $n^C(0)$. Thus, $x(t) \in  \mathcal{O} \forall t\geq T$  for some time $T$;
	if $x_1(0)=13, x_3(0)=5$, and $x_4(0)=64$, then the cooperative line $u^C_2$ is active, driving the solution trajectory to the extreme equilibrium
	$(13,18,5,64)$.
\end{case}
\begin{case}
    For $n^C(0) = 90$, if $x_1(0) = 13$, $x_3(0) \leq 4$, then
    $x(t) = x(0) \in  \mathcal{X}^*\ \forall t$ (note that $x_2(0) \geq 1$ since $n^C(0) \geq 90$).
    If $x_1(0) < 13$ the max active line is a defective line at $n^C(0)$, resulting in
    $x(t) \in  \mathcal{O} \forall t\geq T$  for some time $T$.
    If $x_1(0) = 13$ but $x_3(0) = 5$, then the solution trajectory reaches the extreme equilibrium $(13,18,5,64)$.
\end{case}
\begin{case}
	For $91 \leq n^C(0) < 100$, since $x_2(0)\geq 1$, the solution trajectory reaches the extreme equilibrium $(13,18,5,64)$.
\end{case}

By combining Cases 2 to 4, we obtain the following subset of the basin of attraction of the (union of the) fluctuation set(s) $\O$:
\begin{equation*}
    \left\{x\in\X\,|\, 27\leq n^C(x) \leq 81 \right\}.
\end{equation*}

\section{Approximating the basin of attraction} \label{sec:basin}
Motivated by the example, we establish the following proposition to approximate the basin of attraction of the fluctuation interval $\mathcal{I} = [a,b]$.
In view of Proposition \ref{prop:max}, the max lines at $a$ and $b$ are cooperative and defective, respectively, denoted by $u^C_i$ and $u_k^D$ for some types $i$ and $k$.
If there is an intersection point $p_1$ on $u_i^C$ such that there is no intersection point at any $n^C \in (n^C(p_1), n^C(p_{ij}))$, then define $a' = n^C(p_1)$, otherwise, $a' = 0$.
If there is an intersection point $p_2$ on $u_k^D$ such that there is no intersection point at any $n^C \in (n^C(p_{lk}), n^C(p_{2}))$, then define $b' = n^C(p_2)$, otherwise, $b' = n$.
Define $Q = Q_1\cup Q_2$, where
\begin{align*}
        Q_1 &= \left\{ x\in\X \,|\, n^C(x) \in [a', n^C(p_{ij})], x_i\geq 0 \right\}, \\
        Q_2 &= \left\{ x\in\X \,|\, n^C(x) \in [n^C(p_{lk}),b'], x_k \leq n_k-1 \right\}.
\end{align*}
\begin{proposition} \label{pb}
    If $x(0) \in Q$, then there exists some time $T\geq 0$ such that $n^C(t)\in \I\, \forall t\geq T$.
\end{proposition}

So $Q$ is a subset of the basin of attraction of the union of the fluctuation sets with the corresponding fluctuation interval $\I$.

% Let $p_3$ be intersection points on $u_i^C$ or
% any cooperative lines above $u_i^C$ at $n^C(p_{ij})$,
% $p_4$ be intersection points on $u_k^D$ or
% any defective lines above $u_k^D$ at $n^C(p_{lk})$
% such that there is no other intersection point at $n^C \in [n^C(p_3), n^C(p_{ij})]$
% or $n^C \in [n^C(p_{lk}), n^C(p_{4})]$. Let $\str D(S,p_{ij})$, where $S\in \{C,D\}$, denote the set of agent types whose corresponding S lines are below $p_{ij}$ at $n^C(p_{ij})$.
% \begin{proposition}
% 	For any $x(t)$ with 
% 	\[n^C(x(t)) \in [max(n^C(p_3),1+\sum_{q \in \str D(C,p_{ij})}{n_q}), n^C(p_{ij})]\] or 
% 	\[n^C(x(t)) \in [min(n^C(p_4),1+\sum_{q \in \str T(D,p_{lk})}{n_q}), n^C(p_{lk})]\text{, }\]
% 	we have $x(t) \in  \mathcal{O} \quad \forall t\geq T$  for some time $T$.
% \end{proposition}
% \begin{IEEEproof}
%     Consider a 
% \end{IEEEproof}

\section{convergence analysis for well-known $2\times2$ games} \label{sec:analysis}
After presenting the general results on the asymptotic behavior of imitation dynamics, we now apply them to the three acknowledged $2\times2$ games \cite{riehl2018survey}.  

\subsection*{Coordination}
The payoff matrix of an agent $i$ is called a \emph{coordination payoff matrix} if 
\begin{equation}    \label{coordinationConstraintOnPayoffs}
\textcolor{Blue}{R_i}>\textcolor{RawSienna}{T_i},\ 
\textcolor{Red}{P_i}>\textcolor{Emerald}{S_i}.
\end{equation}
The agent then earns more if she plays the same strategy as her opponent's.
In other words, myopically, she profits by coordinating with her opponent, hence, the name `coordination'. 
The above constraint on the payoffs restrains the cooperative and defective lines of the same type of agents to form a potentially attracting intersection point.
\begin{lemma} \label{coord}
    Consider an exclusive population of agents with coordination payoff matrices.
	For any potentially attracting intersection point $p_{ij}$ of agent types $i$ and $j$ in the corresponding graphical representation, it holds that $i \neq j$.
\end{lemma}

So, potentially attracting intersection points, and hence, fluctuation points can occur only between lines of different agent types.
As a result, we show that it is impossible to have both a left wall and a right wall in the population, forcing the dynamics to equilibrate. 
\begin{proposition} \label{th_coord}
	Every exclusive population of agents with coordination payoff matrices and governed by imitation dynamics will equilibrate.
\end{proposition}

We note that Lemma \ref{coord} and Proposition \ref{th_coord} still hold if \eqref{coordinationConstraintOnPayoffs} is relaxed to $\textcolor{Blue}{R_i} + \textcolor{Red}{P_i} > \textcolor{RawSienna}{T_i} + \textcolor{Emerald}{S_i}$.
This allows for the payoff matrices to be those of an opponent-coordinating game \cite{riehl2018survey}, where $\textcolor{Blue}{R_i} > \textcolor{Emerald}{S_i}$ and $\textcolor{Red}{P_i} > \textcolor{RawSienna}{T_i}$.

%%%%%%%%%%%%%%%
\subsection*{Prisoner's Dilemma}
The payoff matrix of an agent $i$ is called a \emph{prisoner's dilemma payoff matrix} if 
\begin{equation} \label{PDConsdtraintOnPayoffs}
\textcolor{RawSienna}{T_i}> \textcolor{Blue}{R_i},
\textcolor{Red}{P_i}>\textcolor{Emerald}{S_i} \text{.}
\end{equation}
This payoff constraint, results in the following strong restriction on the graphical representation.
\begin{lemma} \label{prison}
    Given an exclusive population of agents with prisoner's dilemma payoff matrices,
    a defective line is above all other lines (at every point $n^C$) in the graphical representation.
\end{lemma}

Similarly, the following result can be shown.
\begin{lemma} \label{prison2}
    Consider an exclusive population of agents with prisoner's dilemma payoff matrices.
	For any potentially attracting intersection points $p_{ij}$ of agent types $i$ and $j$ in the corresponding graphical representation, it holds that $i \neq j$.
\end{lemma}

So similar to the coordination case, Lemma \ref{prison2} implies that the dynamics converge to an equilibrium. 
However, thanks to Lemma \ref{prison}, we can provide a simpler proof here.
\begin{proposition} \label{p99}
	Every exclusive population of agents with prisoner's dilemma payoff matrices and governed by imitation dynamics will equilibrate. 
\end{proposition}

%%%%%%
\subsection*{Coordination and Prisoner's Dilemma}
Because of Lemmas \ref{coord} and \ref{prison2}, we can combine the results of the previous two subsections as follows.
\begin{theorem} \label{th6}
    Every exclusive population of agents with either prisoner's dilemma or coordination payoff matrices and governed by imitation dynamics will equilibrate. 
\end{theorem}

\subsection*{Anticoordination}
The payoff matrix of an agent $i$ is called a \emph{anticoordination payoff matrix} if 
\begin{equation} \label{SDConsdtraintOnPayoffs}
\textcolor{RawSienna}{T_i}> \textcolor{Blue}{R_i}
\text{, }
\textcolor{Emerald}{S_i}>\textcolor{Red}{P_i} \text{.}
\end{equation}
The agent then earns more if she plays the opposite strategy of her opponent's.
In other words, myopically, she profits by anti-coordinating with her opponent, hence, the name `anticoordination'. 
Note that games such as \emph{snowdrift} and \emph{chicken game} have payoff matrices that also follows the above constraint. 
Unlike with the coordination and prisoner's dilemma cases, the above constraint guarantees a potentially attractive intersection point formed by the two lines of every agent type.
\begin{lemma} \label{le10}
	Corresponding to every agent $i$ with an anticoordination payoff matrix, 
	there exists an potentially attracting intersection point $p_{ii}$.
\end{lemma}

\begin{figure}  
	\centering
	\begin{subfigure}[b]{0.48\linewidth}
	\begin{tikzpicture}[scale=0.50]
		\begin{axis}[
			standard,
			xlabel=$n^C$,
			xtick={50},
			ylabel=$u$,
			legend pos=outer north east,
			legend style={font=\small},
			every axis y label/.style={
    		at={(ticklabel* cs:1.02)},
			anchor=east,
			},
			every axis x label/.style={
    		at={(ticklabel* cs:1.1)},
    		anchor=east
}
		]
		\addplot [
			name path=Q1,
			domain=0:100, 
			samples=10, 
			color=blue,
			] coordinates
			{(0,2)
			(100,0)};
		\addlegendentry{$u_i^C$}

		\addplot [
			name path=Q2,
			domain=0:100, 
			samples=10, 
			color=blue,
			style=dashed
			] coordinates
			{(0,0)
			(100,2)};
		\addlegendentry{$u_i^D$}
		%Here the blue parabloa is defined
		\path [name intersections={of=Q1 and Q2,by=Q}];
		\node [circle,fill=black,inner sep=1.5pt,label=90:$p_{ii}$] at (Q) {};

		\draw [gray, dashed] (Q) -- (50,0);
		
		\end{axis}
	\end{tikzpicture}
	\caption{}
	\end{subfigure}\hfill
	\begin{subfigure}[b]{0.48\linewidth}
		\begin{tikzpicture}[scale=0.50]
			\begin{axis}[
				standard,
				xlabel=$n^C$,
				xtick={50.5},
				ylabel=$u$,
				legend pos=outer north east,
				legend style={font=\small},
				every axis x label/.style={
    		at={(ticklabel* cs:1.1)},
    		anchor=east
},
				every axis y label/.style={
				at={(ticklabel* cs:1.02)},
				anchor=east,
	}
			]
			\addplot [
				name path=C1,
				domain=0:100, 
				samples=10, 
				color=blue,
				] coordinates
				{(0,2)
				(100,0.04)};
			\addlegendentry{$u_1^C$}
	
			\addplot [
				name path=C2,
				domain=0:100, 
				samples=10, 
				color=blue,
				style=dashed
				] coordinates
				{(0,0)
				(100,2)};
			\addlegendentry{$u_1^D$}
			%Here the blue parabloa is defined
			\path [name intersections={of=C1 and C2,by=C}];
			\node [circle,fill=black,inner sep=1.5pt,label=90:$p_{11}$] at (C) {};
	
			\draw [gray, dashed] (C) -- (50.5,0);
			\end{axis}
		\end{tikzpicture}
		\caption{}
	\end{subfigure}
	%\captionsetup{font=tiny}
	\caption{
	All agents in both populations share the same anticoordination payoff matrix, and hence, belong to the same type $1$.
	\textbf{a)} $p_{11}$ is an integer intersection point.
	Unless the solution trajectory starts from either of the extreme equilibria, it reaches the equilibrium $x^*=(n^C(p_{11}))$.
	\textbf{b)} $p_{11}$ is a fluctuation point.
	Unless the solution trajectory starts from either of the extreme equilibria, it ends up fluctuating between $x^1 = (50)$ and $x^2 = (51)$ forever.
    }
	\label{fig7-1}
\end{figure}
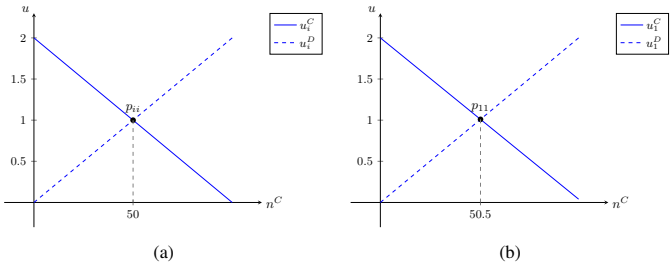

The existence of walls, therefore, may not be excluded. 
Indeed, both possible asymptotic behaviors of the imitation dynamics are covered by single-type populations of agents with the same anticoordination payoff matrix (Fig.~\ref{fig7-1}).
Non-convergence to equilibrium states in imitation dynamics is, hence, due to the existence of agents with anti-coordination payoff matrices as any population of agents with the other two types of payoff matrices equilibrate. 

\section{Concluding remarks} \label{sec:concludingRemarks}
Finite well-mixed heterogeneous populations of agents who asynchronously imitate the one with the highest payoff may either reach an equilibrium or undergo perpetual fluctuations in a minimal positively invariant set in the long run.
In addition to finding the equilibria of the dynamics and determining their stability, for the first time, we have characterized all positively invariant sets and approximated their basin of attractions. 
%We have found that the non-converging behaviors are caused by anticoordinators. 
Our results apply to social contexts where individuals are likely to take the same action as the most successful individual, for example, breaking social norms among teenagers, blindly investing on what successful firms are, and volunteering to establish a social service. 
For a population of imitators that never settles on a state, our analysis suggests the existence of a group of individuals who have higher payoffs when they cooperate if the cooperation level is low and defect if the cooperation level is high.
%This together with the characterization of the fluctuation intervals and equilibria, pave the way for controlling the total number of cooperators in the population as well as shaping the length of the oscillations. 

\emph{
Extensions to several aspects of our model remain as future work:
\emph{(i)} increasing the number of available strategies;
\emph{(ii)} relaxing the well-mixed assumption and considering a structured population; 
\emph{(iii)} relaxing the matrix-game assumption, and allowing the
utilities to be a non-linear function of the payoffs. 
}

%We have also performed stability analysis, showing that in a heterogeneous population the states where everyone is satisfied with their action is unstable, except for the states where everyone takes the same action. 
%These results accord with our intuition on the irrational nature of imitation dynamics, and illustrate the observations in the literature.

\appendix
% The following result is straightforward.
\begin{lemma} \label{lem2} %was 13
	A cooperative type-$i$ agent and a defective type-$j$ agent ($i$ and $j$ are not necessarily different) have the same payoff at $n^C \in [0,n]$ if and only if $(a_i-c_j)n^C = d_j-b_i$. 
\end{lemma}
\begin{IEEEproof}
	\begin{equation*} 
		n^C = \frac{d_j-b_i}{a_i-c_j} \xLeftrightarrow{\eqref{eq3}}u_i^C(n^C)=u_j^D(n^C)\text{.}
	\end{equation*}
\end{IEEEproof}

\begin{IEEEproof}[Proof of Theorem \ref{th1}]
    (Sufficiency)
    In Case 1), since for both states, agents are either all cooperating or defecting, then according to \eqref{updateRule}, no agent can switch her strategy in the next time step for all time. Therefore, we have $x\in\mathcal{X}^*$.
    Now we show the proof for Case 2).
    First, note that since $(a_i-c_j)n^C(x) = d_j-b_i$,
	it follows from Lemma \ref{lem2} in the Appendix that there are a  type-$i$ cooperator and a type-$j$ defector having the same payoff at $n^C(x)$.
    Now we prove by contradiction. 
    Assume on the contrary, $x\notin \mathcal{X}^*$. Then by definition, there exists an agent of type $k \notin \{i,j\}$ such that either $u^C_k(n^C(x))>u^C_i(n^C(x)) = u^D_j(n^C(x))$ with $x_k > 0$ or $u^D_k(n^C(x))>u^C_i(n^C(x)) = u^D_j(n^C(x))$ with $x_k < n_k$.
    Both cases contradict the last condition in the lemma. Therefore, we have $x\in\mathcal{X}^*$.
    
	(Necessity) 
	Consider an equilibrium state $x \in \mathcal{X}^*$. 
	If $n^C(x)=0$ or $n$, then $x=(0,\ldots,0)$ or $(n_1,\ldots,n_m)$. If $n^C(x)\in(0,n)$, then there are both cooperating and defecting agents in the population. 
	Since they never change their strategies upon activation, according to (1) there must exist a cooperating agent of say type-$i$ and a defecting agent of say type-$j$, both earning the highest payoffs. 
	This implies that $x_i \neq 0$, $x_j \neq n_j$, and also $(a_i-c_j)n^C(x) = d_j-b_i$ in view of Lemma \ref{lem2}. 
	Moreover, since $u_i^C$ and $u_j^D$ are both the greatest utilities at $n^C(x)$, if there is a type-$k$ agent with $u_k^C(n^C) > u_i^C(n^C)$, i.e., $a_kn^C+b_k>a_in^C+b_i$, then it must be defecting, i.e., $x_k = 0$, and if there is a type-$k$ agent with $u_k^D(n^C) < u_i^D(n^C)$, i.e., $c_kn^C+d_k>c_in^C+d_i$, then it must be cooperating, i.e., $x_k = n_k$. 
	This completes the proof.
\end{IEEEproof}

\begin{IEEEproof}[Proof of Proposition \ref{prop11}]
	We prove by showing that Proposition \ref{prop11} is a graphical interpretation of Theorem \ref{th1}. 
	Since $n^C((0,\ldots,0)) = 0$ and $n^C(n_1,\ldots,n_m) = n$, we know that the first case in Theorem \ref{th1} is equivalent to the first case in Proposition \ref{prop11}. 
	On the other hand, from \eqref{eq3} and Lemma \ref{lem2}, we know that the second case in Theorem \ref{th1} is equivalent to the case where cooperating agents of type $i$ and defecting agents of type $j$ are both having the highest payoff. 
	This is also equivalent to the second case of the Proposition.
\end{IEEEproof}

%The following proof is for Lemma \ref{lem4}.
\begin{IEEEproof}[Proof of Lemma \ref{lem4}]
	First, $x(t)$ does belong to a positively invariant set for all $t\geq 0$ as $x(t)\in \mathcal{X}$ and $\X$ is positively invariant.
	Moreover, by definition, $\mathcal{X}$ has one or more minimal positively invariant subsets $\mathcal{X}_1,\ldots,\mathcal{X}_p, p\in\Z_{\geq0}$. 
	By contradiction, we prove that $x$ belongs to one of them for some $T>0$. 
	Assume on the contrary that for all $t>0$, $x(t)$ is not contained in any of them. 
	Thus, for all $t> 0$, $x(t)$ is confined to the subset $\mathcal{S}\subseteq \mathcal{X}-\cup_{i=1}^{p} \mathcal{X}_i$, which is nonempty. 
	Being finite, $\mathcal{S}$ has a nonempty subset $\mathcal{S}'$ of states, each of which is visited by $x$ infinitely many times. 
	Now if there exists a trajectory $y(t)$ that starts from $\mathcal{S}'$ but leaves it at some time $t'>t$, then $x(t)$ cannot intersect with $y(t)$ since, otherwise, $x(t)$ would also follow $y(t)$ and leave $\mathcal{S'}$ at some time due to the randomness of the activation sequence. 
	Exclude the states of all such outgoing trajectories $y(t)$ from $\mathcal{S'}$ to obtain the set $\mathcal{S''}$. 
	The set $\mathcal{S''}$ is positively invariant since there are no trajectories leaving it, and nonempty since $x(t)\in\mathcal{S''}$ for all $t\geq0$.
	Hence, $\mathcal{S''}$ contains a minimal positively invariant subset $\mathcal{S'''}$ that itself is a minimal positively invariant subset of $\mathcal{X}$. 
	However, $\mathcal{S'''}$ does not equal any of $\mathcal{X}_i$s, which is a contradiction, completing the proof.
\end{IEEEproof}

\begin{IEEEproof}[Proof of Proposition \ref{prop:2}]
    (Sufficiency) 
    From Lemma \ref{lem4} we know that there exists a minimal positively invariant set $\O$ such that $x(T) \in \O$ for some $T\geq 0$.     
    We claim that $x(t)$ never reaches an equilibrium. 
    Assume on the contrary that there exists some $x'' \in \mathcal O$ such that $x(T) \neq x''$ for any $T\geq 0$. Then the set $\mathcal O \setminus \{x''\}$ is a positively invariant set and $\mathcal O \setminus \{x''\} \subset \mathcal O$, which contradicts the minimality of $\mathcal O$.
    Since $x(t)$ never reaches an equilibrium, we know that $\O$ does not contain an equilibrium state, otherwise $x(t)$ can reach it under some activation sequence. 
    Thus, the set $\O$ is a non-singleton minimal positively invariant set such that $x(T)\in \O$ for some $T \geq 0$. \\
	(Necessity) 
	Since $x(T)\in \O$ for some $T \geq 0$, where $\O$ is a non-singleton minimal positively invariant set. By the definition of non-singleton minimal positively invariant sets, we know that $x(t)$ never reaches an equilibrium.
\end{IEEEproof}

\begin{IEEEproof}[Proof of Lemma \ref{lem5}]
    We prove this by contradiction. Assume on the contrary that there exists a state $y \in \O$ such that there are only finite many time instants $\{t_1, \ldots, t_s\}$ at which $x(t_i) = y, i \in \{t_1, \ldots, t_s\}$. By definition, this implies that the set $\O \setminus y$ is a positively invariant set. This contradicts the fact that $\O$ is minimal.
\end{IEEEproof}

\begin{IEEEproof}[Proof of Lemma \ref{lem6}]
    We only show the result for all $x\in\mathcal{O}$ with $n^C(x) = a$. 
    We prove by contradiction.
    Assume on the contrary that there exists a state $x' \in \mathcal{O}$ with $n^C(x') = a$ such that among the lines active by $x'$, the max at $a$ is a defective line. 
    Consider some $x(0)\in\mathcal{O}$, and let $T_1$ be some time when $x(t)$ equals $x'$. 
    If the active agent at time $T_1$ is a defector, the state remains unchanged. 
    However, in view of Lemma \ref{lem5}, the state $x(t)$ does not remain at $x'$ forever and has to visit the other states in $\mathcal{O}$.
    Therefore, a cooperator will become active at or after $T_1$. 
    Let $T_2$ be the first time this happens. 
    Then, 
    \begin{equation*}
        n^C(x(T_2+1)) = a - 1
        \Rightarrow
        x(T_2+1) \notin \mathcal{O}.
    \end{equation*}
    which is a contradiction, completing the proof.
\end{IEEEproof}

\begin{IEEEproof}[Proof of Proposition \ref{prop:max}]
    We only show the result for the max line at $a$. 
    We prove by contradiction.
    Assume on the contrary that the max line at $a$ is a defective line $u^D_k$ for some type $k$. 
    From the definition of fluctuation sets we know that there exists a state $y \in \O$ with $n^C(y) = a+1$ such that the max active-by-$y$ line is a defective line. 
    Consider the initial condition $x(0) = y$.
    If $y_k \geq 1$, i.e., at least one type-$k$ agent is a cooperator at $y$, then under the activation sequence where a type-$k$ cooperator is active at $t=0$ and another cooperator is active at $t=1$, we obtain $n^C(1) = a$ and $u^D_k$ is active by $x(1)$, resulting in $n^C(2) = a-1$.
    Hence, $x(2) \not\in\O$, which is a contradiction.
    Note that such an activation sequence exists since $n^C(y) = a+1\geq2$.
    If $y_k = 0$, i.e., all type-$k$ agents are defectors at $y$, then $u^D_k$ is already active by $y$. 
    Thus, under the activation sequence where two different cooperators are active at $t=0$ and $t=1$, we obtain $n^C(1) = a$ and $n^C(2) = a-1$, resulting in, again, a contradiction. 
\end{IEEEproof}

\begin{IEEEproof}[Proof of Lemma \ref{fluct_pt}]
    Let $\O$ be a fluctuation set with the corresponding interval $\I$.
    Because of the minimality of $\O$, there exists a state $y\in\O$ with $n^C(y) = a+1$ such that among the lines active-by-$y$, the max at $a+1$ is defective.
    Consider the initial state $x(0) = y$. 
    Under the activation sequence where a cooperator is active at time $t=0$, we obtain the state $x(1)$ with $n^C(x(1)) = a$. 
    On the other hand, in view of Lemma \ref{lem6}, the max line among those active by $x(1)$ is cooperative.
    Hence, the cooperative line at $a$ intersects with the defective line at $a+1$ in the interval $(a,a+1)$, resulting in a non-integer potentially attracting intersection point, which is a fluctuation point since $y\in\X_+(x(1))$.
    Similarly, there must exist a fluctuation point in $(b-1,b)$, completing the proof.
\end{IEEEproof}

\begin{IEEEproof}[Proof of Lemma \ref{wall}]
    %We only prove the case with $|\mathcal{R}\setminus \mathcal{S}| < a$. 
    (Necessity) 
    Since $[a,a+1]$ contains a fluctuation point, there exists a state $y \in \X$ with $n^C(y) = a$ such that the max active-by-$y$ line at $a$ is cooperative, and there exists a state $y_{+}\in\X_{+1}(y)$ such that the max active-by-$y_+$ line at $a+1$ is defective.
    It is impossible to have all defectors at $y$ being in $\S$ since then for any state $y_{+}\in\X_{+1}(y)$, one of the cooperative lines above $u^D_s$ at $a+1$ is active by $y_+$, making the max active-by-$y+$ line cooperative.
    % Either of the following two cases holds.    
    % \begin{case}
    %     \emph{All defectors at $y$ are in $\S$. }
    %     Then for any state $y_{+}\in\X_{+1}(y)$, one of the cooperative lines above $u^D_s$ at $a+1$ is active by $y_+$.
    %     Hence, the max active-by-$y+$ line is a cooperative, which makes this case infeasible.
        % Then we have
        % \emph{Statement 1: all agents in $\mathcal{R}\setminus \mathcal{S}$ are cooperators.}
        % Now, we show by contradiction that \emph{Statement 2: the set $\mathcal{R}\setminus \mathcal{S}$ does not contain all of the cooperators in the population}. 
        % Assume the contrary.
        % Then $u^D_r$ and all cooperative lines above $u^D_r$ at $a$ are inactive. 
        % Now consider the initial state $x(0) = y$. 
        % Under the activation sub-sequence $(q^1,r^1)$, where $q^1$ is any defector, and $r^1$ is a type-$r$ cooperator, we have the max active-by-$x(2)$ line to be $u^D_r$, a defective line.
        % However, if a cooperator is active at $t = 2$, we obtain $n^C(x(3)) < a$, contradicting the interval $[a,a+1]$ being a wall.
        % Hence, Statement 2 is in force.
        % Therefore, the two statements together with the fact that the total number of cooperators at $y$ is $a$, result in $|\mathcal{R}\setminus \mathcal{S}| < a$.
        % This completes the proof of this case.
    % \end{case}
    % \begin{case}
        Thus, at least one defector at $y$, say agent $d$, is not in $\S$. 
        Now, first, we prove by contradiction the existence of a state $z\in\X$ with $n^C(z) = a+1$ such that the max active-by-$z$ line at $a+1$ is the defective line $u^D_s$. 
        Assume on the contrary that there is no such state.
        Since the interval $[a,a+1]$ contains a fluctuation point, there exists $v\in\X$ with $n^C(v) = a+1$ such that the max active-by-$v$ line is a defective line other than $u^D_s$.
        Since the max defective line at $a+1$ is $u^D_s$, and $u^D_s$ is inactive by $v$, there exists a type-$s$ cooperator at the state $v$, denoted by $s^1$.
        Then starting from the initial state $x(0) = v$ and under the activation sub-sequence $(s^1,d)$, we have $x(2) = v$ with $n^C(v) = a+1$ such that the max active-by-$v$ line is $u^D_s$, resulting in a contradiction. 

        Next, we complete the proof by contradiction.
        Assume on the contrary that $|\mathcal{R}\setminus \mathcal{S}| \geq a$. 
        We construct an activation sequence that drives the initial state $x(0) = z$, to a state with the number of cooperators less than $a$, i.e., $n^C < a$.
        Let $\mathcal{C} = \{c^1, c^2, \ldots, c^{k_c}\}$ denote all of the cooperators at $z$ who are not in $\mathcal{R}$ and $\mathcal{D} = \{d^1, d^2, \ldots, d^{k_d}\}$ denote all of the defectors at $z$ who are in $\mathcal{R}\setminus \mathcal{S}$. 
        Since no cooperative line that is above $u^D_s$ at $a+1$ is active by $z$, there is no cooperator at $z$ that belongs to $\mathcal{S}$. 
        Thus, there are $a+1-k_c$ number of cooperators in $\mathcal{R}\setminus \mathcal{S}$. 
        Then the number of defectors in $\mathcal{R}\setminus \mathcal{S}$ is 
        \begin{equation*}
            k_d = |\mathcal{R}\setminus \mathcal{S}| - (a+1-k_c) \text{.}
        \end{equation*}
        Since $|\mathcal{R}\setminus \mathcal{S}| \geq a$, we have $k_c \leq k_d+1$.
        So the following activation sequence is well defined:
        \begin{equation*}
            (c^1, d^1, c^2, d^2,\ldots,  d^{k_c},c^{k_c}) \text{.}
        \end{equation*}
        Consider the initial state $x(0) = z$.
        Upon the activation of each agent $c^i, i=1,\ldots, k_c,$ in this sequence, she switches to defection as the max active-by-$z$ line is $u^D_s$ at $a+1$, it never becomes inactive, and no cooperation line above $u^D_s$ at $a+1$ becomes active by any of the $d^i, i=1,\ldots, k_c,$ since they are not in $\S$. 
        By the end of this sequence, i.e., when $c^{k_c}$ switches to defection, we have $n^C(x(T+1)) = a$, were $T = 2k_c$.
        Moreover, there is no cooperator left at $x(T+1)$ that is not in $\mathcal{R}$, and hence, $u^D_r$ becomes the max active-by-$x(T+1)$ line. 
        Hence, if any cooperator becomes active at $T+1$, we obtain $n^C(x(T+2)) < a$, which is a contradiction, completing the proof of this case.
    % \end{case}
    
    (Sufficiency) 
    We prove by contradiction.
    Assume on the contrary that there exists an initial state $x(0)$ with $n^C(x(0)) = a+1$, such that under some activation sequence, $n^C(x(1)) = a$ and $n^C(x(2)) = a-1$. 
    This implies that the max active lines by $x(0)$ and $x(1)$ are both defective at $n^C = a+1$ and $n^C = a$. 
    Hence, all agents in $\S$ are defectors at $x(0)$, and remain so at $x(1)$. 
    So in view of $|\mathcal{R} \setminus \mathcal{S}| < a$, there is a cooperator at $x(1)$ who is not in $\mathcal{R}$.
    Thus, there is a cooperative line above $u^D_r$ at $a$ being active by $x(1)$, implying that the max active-by-$x(1)$ line at $a$ cooperative.
    This contradicts our earlier conclusion, which completes the proof.
\end{IEEEproof}

\begin{IEEEproof}[Proof of Theorem \ref{endinter}]
   The proof is similar to that of lemma \ref{wall}. 
   We only prove that $[a,a+1]$ is a left wall.
   Let $\O$ be the corresponding fluctuation set of $\I$. 
   The minimality of $\O$ yields the existence of $z \in \O$ with $n^C(z) = a+1$ such that the max active-by-$z$ line at $a+1$ is defective. 
   Moreover, $[a,a+1]$ contains a fluctuation point thanks to Lemma \ref{fluct_pt}.
   So the rest of the proof can be done similarly to that of Lemma \ref{wall}.
\end{IEEEproof}

\begin{IEEEproof}[Proof of Corollary \ref{constraint}]
From Theorem \ref{endinter} we know that $[a,a+1]$ is a left wall and $[b-1,b]$ is a right wall. By the definition of walls, we know that for any initial condition $x(0)$ with $n^C(x(0)) \in [a+1,b-1]$, we have $n^C(x(t)) \in \I ~\forall t>0$.
\end{IEEEproof}

\begin{IEEEproof}[Proof of Corollary \ref{c2}]
If there is no interval $\I = [a,b],a,b\in\{1,\ldots,n-1\}$ such that $[a,a+1]$ is a left and $[b-1,b]$ is a right wall, then according to Theorem \ref{endinter} there is no fluctuation interval, which by definition implies that there is no fluctuation set, i.e. non-singleton minimal positively invariant set. From Proposition \ref{prop:2} we know that this implies that $x(t)$ reaches an equilibrium.
\end{IEEEproof}

\begin{IEEEproof}[Proof of Proposition \ref{th2}]
    Since $[a,a+1]$ and $[b-1,b]$ are left and right walls, in view of Corollary \ref{constraint}, for any $x(0)\in\mathcal{Q}$ in \eqref{eq:Q}, we have $x(t) \in \mathcal{Q}\ \forall t\geq 0$.
    On the other hand, according to Lemma \ref{lem4}, $x(t)$ will reach a non-singleton minimal positively invariant set $\O$ since $\I$ does not contain any equilibrium. 
    Let $\I'$ be the fluctuation interval corresponding to $\O$. 
    It must hold that $\I'\not\subset\I$ as, otherwise, $\I$ contains another wall due to Theorem \ref{endinter}.
    Moreover, $\I' \cap ([0,n]-\I)= \emptyset$ because of the two walls.
    Hence, $\I' = \I$, implying that $\I$ is the corresponding interval of a fluctuation set, completing the proof.
\end{IEEEproof}

\begin{IEEEproof}[Proof of Theorem \ref{th2_pts}]
By Lemma \ref{wall} and \ref{lem:rightWall}, condtion $2)$ implies that $[\floor{n^C(p_{ij})},\floor{n^C(p_{ij})}+1]$ is a left wall and $[\ceil{n^C(p_{lk})}-1,\ceil{n^C(p_{lk})}]$ is a right wall. By condition $1)$ the definition of walls, we know that there exists a positively invariant set $\O$ such that $\mathcal{I} = [\min_{x\in\mathcal{O}} n^C(x),\max_{x\in\mathcal{O}} n^C(x)]$. From condtion $3)$, we know that the set $\O$ is minimal. From condtion $4)$, we know that the set is non-single. Therefore, the set $\O$ is a fluctuation set and the interval $\I$ is a fluctuation interval.
\end{IEEEproof}

\begin{IEEEproof}[Proof of Proposition \ref{p5}]
    The sufficient result is trivial, so we proceed to the necessity.
    If $s$ is stable, then for the initial condition $x(0) = (x_1-1) \in \mathcal{B}_2(s)$, it must hold that $x(t) \in \mathcal{B}_2(s) = \{(x_1-1),(x_1), (x_1+1)\}\ \forall t\geq0$. 
    Hence, $x(1) \in \{(x_1-1),(x_1)\}$, implying that the max active-by-$x(0)$ line at $x_1$ cannot be defective, yielding $u^C_1(x_1 - 1) \geq u^D_1(x_1 - 1)$. 
    If $s$ is asymptotically stable, then the equality case may no longer hold since it results in $(x_1-1)$ being an equilibrium, which prevents the solution trajectory from returning to $s$.
    The case with $x(0) = (x_1+1)$ is similar and completes the proof.
\end{IEEEproof}

\begin{IEEEproof}[Proof of Proposition \ref{p6}]
    We only show the proof for $s = (0,\ldots,0)$.
    
    (Necessity)
    If $s = (0,\ldots,0)$ is unstable, then there exists a ball $\mathcal{B}_\epsilon(s)$ such that for any $\mathcal{B}_\delta(s)$ there exists an initial condition $x(0)\in\mathcal{B}_\delta(s)$, such that $x(t)\notin \mathcal{B}_\epsilon(s)$ for some $t \geq 1$ and under some activation sequence. 
    Since $\mathcal{B}_2(s) \subseteq \mathcal{B}_\epsilon(s)$ and $\delta\geq 2$, this implies the existence of an  $x(0)\in\mathcal{B}_2(s)$, such that $x(t)\notin \mathcal{B}_2(s)$ for some $t \geq 1$ and under some activation sequence. 
    Thus, for some $x \in \mathcal{X}$ with $n^C(x) = 1$, the max active-by-$x$ line at $n^C = 1$ is a cooperative line of, say, type $i$. 
    Hence, $x = (0,\ldots,0,x_i = 1,0,\ldots,0)$, and $u_i^C(1)$ must be higher than all active defective lines at $n^C = 1$, which are all types $j\neq i$, as well as type $i$ if $n_i >1$.
    This completes the proof.
    
    (Sufficiency)
    If there exists a type-$i$ agent with $n_i > 1$ such that $u_i^C(1)>u_j^D(1)$ for every type-$j$ agent, then for $x(0) = (0,\ldots,x_i = 1,\ldots,0)$, the max active line at $n^C = 1$ is cooperative. 
    Hence, upon the activation of a defector at time $t=0$, $x(1) \notin \mathcal{B}_2(s)$, implying the instability of $s$.
    The proof is similar for $n_i = 1$.
\end{IEEEproof}

\begin{IEEEproof}[Proof of Theorem \ref{st1}]
    Consider the extreme equilibrium state $s = (0,\ldots,0)$. 
    First, we prove the stability result.
    (Necessity)
    Without loss of generality, we show that for every $x(0)\in\mathcal{B}_2(s)$, we have $x(t)\in \mathcal{B}_2(s)\ \forall t \geq 0$ under any activation sequence. Since at least one max line is defective at $1$, then if $n^C(x(t)) = 1$, either the highest-earning strategy is defection or there is an equilibrium state. In both case, we have $x(t)\in \mathcal{B}_2(s)\ \forall t \geq 0$ under any activation sequence.
    
    (Sufficiency)
    The result simply follows from the fact that if $s$ is stable, then for every $x(0)\in\mathcal{B}_2(s)$, we have $x(t)\in \mathcal{B}_2(s)\ \forall t \geq 0$ under any activation sequence. This implies that there is at least one defective max line at $n^C = 1$.
    This also proves the asymptotic stability part. 
\end{IEEEproof}

\begin{IEEEproof}[Proof of Theorem \ref{st2}]
    First, we prove the stability result.
    (Necessity)
    We prove by contradiction.
    Assume on the contrary that there exists a non-equilibrium state $y\in \X$ with $||x^*-y||=1$.
    Without loss of generality, let $n^C(y) = n^C(x^*) + 1$ with $y_i = x^*_i+1$ for some agent type $i$.
    If the max active-by-$y$ line is a cooperative line of type $k$, then for $x(0) = y \in \mathcal{B}_2(x^*)$, we have $x(1) \notin \mathcal{B}_2(x^*)$ if a defector becomes active at time $t=0$, which is a contradiction. 
    Note that a defector exists at $y$ since, otherwise, $y$ is an equilibrium.
    If the max active-by-$y$ line is a defective line of type $k$, but there exists an agent type $j \neq  i$ such that $y_j>0$, then for $x(0) = y\in \mathcal{B}_2(x^*)$, we have $x(1) \notin \mathcal{B}_2(x^*)$ if a cooperating agent of type-$j$ becomes active at time $t=0$, which leads to a contradiction.
    So $y_j = 0\ \forall j\neq i$.
    However, in view of $\|y-x^*\| = 1$, this implies $x_j = 0\ \forall j \neq i$.
    Now the same must hold for any other state $y\in \X$ satisfying $||x^*-y||=1$, but with $y_p = x^*_p+1$ for some type $p\neq i$.
    Hence, 
    $x_j = 0\ \forall j \neq p$ as well.
    Therefore, $x_j = 0\ \forall j$, resulting in $x = (0,\ldots,0)$, which is impossible.
    %The proof when $n^C(y) = n^C(x^*) - 1$ is similar.
    
    (Sufficiency)
    The result simply follows from the fact that starting from an initial condition with distance $1$ from $x^*$, the solution trajectory remains there forever.
    This also proves the asymptotic stability part. 
\end{IEEEproof}

\begin{IEEEproof}[Proof of Proposition \ref{pb}]
    There is no defective line above $u^C_i$ at any $n^C \in [a, n^C(p_{ij})]$ as there is no intersection point at any $n^C \in (a, n^C(p_{ij}))$. 
    Hence, for $x(0)\in Q_1$, since $x_i(0) \geq 1$, the max active-by-$x(t)$ line is cooperative at all $n^C(x(t)) \in [a, n^C(p_{ij})]$, which drives $n^C(x(t))$ to the fluctuation interval $\I$.
    The proof is similar for $x(0) \in Q_2$.
\end{IEEEproof}

\begin{IEEEproof}[Proof of Lemma \ref{coord}]
    It suffices to show that for any potentially attracting intersection point $p_{ij}$, $i\neq j$.
    Assume on the contrary that there exists a potentially attracting intersection point $p_{ii}$ that is intersected by $u^C_i$ and $u^D_i$ for some agent type $i$.
	From \eqref{eq3}, the slope of $u_i^C$ is $a_i = \textcolor{Blue}{R_i} - \textcolor{Emerald}{S_i}$ and 
	the slope of $u^D_i$ is $c_i = \textcolor{RawSienna}{T_i} - \textcolor{Red}{P_i}$.
	Hence, $a_i>c_i$ according to \eqref{coordinationConstraintOnPayoffs}.
	Therefore, 
    \begin{equation*}
    \begin{split}
       a_in^C(p_{ii}) + b_i & = c_in^C(p_{ii})+d_i \\
       %\Rightarrow a_in^C(p_{ii}) + b_i - a_i & < c_in^C(p_{ii})+d_i - c_i \\
       \Rightarrow a_i(n^C(p_{ii})-1) + b_i & < c_i(n^C(p_{ii})-1)+d_i ,
    \end{split}
    \end{equation*}
	which contradicts the definition of potentially attracting intersection points. 
\end{IEEEproof}

\begin{IEEEproof}[Proof of Proposition \ref{th_coord}]
    Thanks to Corollary \ref{c2}, it suffices to show that there is no fluctuation interval in the population in the view of \ref{c2}.
    We prove by contradiction. 
    Assume on the contrary that there is a fluctuation interval $\I = [a,b], a,b\in\{1,\ldots, n-1\}$ such that $[a,a+1]$ is a left and $[b-1,b]$ is a right wall.
    Thus, from Lemmas \ref{wall} and \ref{lem:rightWall}, we have $|\mathcal{R} \setminus \mathcal{S}| < a$ at $[a,a+1]$, and $|\mathcal{K} \setminus \mathcal{L}| < n-b$ at $[b-1,b]$.
    Let $\mathcal{V}$ denote the set of agents that are not in $\mathcal{R}\setminus \mathcal{S}$ at $[a,a+1]$, which is the set of agents whose cooperative lines are above $u^D_r$ at $a$ and agents whose cooperative lines are below $u^D_r$ at $a$ but above $u^D_s$ at $a+1$. 
    Let $\mathcal{W}$ denote the set of agents that are not in $\mathcal{K}\setminus \mathcal{L}$ at $[b-1,b]$, which is the set of agents whose defective lines are above $u^C_k$ at $b$ and agents whose defective lines are below $u^C_k$ at $b$ but above $u^C_l$ at $b-1$.
    Then we have 
    \begin{equation*}
        \begin{split}
            |\mathcal{V}|+|\mathcal{W}| &= (n-|\mathcal{R} \setminus \mathcal{S}|) + (n - |\mathcal{K} \setminus \mathcal{L}| ) \\
            &> n-a+n-(n-b) = n+(b-a) .
        \end{split}
    \end{equation*}
    Moreover, the cooperative lines of the agents in $\mathcal{V}$ must be below or equal to $u^C_l$ at $b-1$ and $u^C_k$ at $b$ since $u^C_l$ and $u^C_k$ are the max cooperative lines at $b-1$ and $b$ respectively. 
    For the same reason, the defective lines of the agents in $\mathcal{W}$ must be below or equal to $u^D_r$ at $a$ and $u^D_s$ at $a+1$. In other words, we have
    \begin{align*}
        &u^C_i(a)>u^D_j(a)~or~u^C_i(a+1)>u^D_j(a+1)  ~\forall i\in \mathcal{V}, \forall j \in \mathcal{W} \\
        &u^C_i(b-1)<u^D_j(b-1)~or~u^C_i(b)<u^D_j(b)  ~\forall i\in \mathcal{V}, \forall j \in \mathcal{W}.
    \end{align*}
    This implies that the cooperative lines of the agents in $\mathcal{V}$ and defective lines of the agents in $\mathcal{W}$ must intersect, resulting in potentially attracting intersection points. 
    Thus, according to Lemma \ref{coord}, we know that one agent cannot be in both $\mathcal{V}$ and $\mathcal{W}$. In other words, we have $\mathcal{V} \cap \mathcal{W} = \emptyset$. This results in $|\mathcal{V}|+|\mathcal{W}| \leq n$, which is a contradiction.
\end{IEEEproof}

\begin{IEEEproof}[Proof of Lemma \ref{prison}]
    According to \eqref{eq3} and \eqref{PDConsdtraintOnPayoffs},  
	\begin{equation*}
		%\begin{aligned}
			u_i^C-u_i^D=(\textcolor{Blue}{R_i}-\textcolor{RawSienna}{T_i})n^C+(\textcolor{Emerald}{S_i}-\textcolor{Red}{P_i})(n-n^C)<0,
		%\end{aligned}
	\end{equation*}
	implying that 
	$u_i^D(n^C)>u_i^C(n^C)\ \forall n^C\in\{0,\ldots,n\}$.
	This leads to the proof.
\end{IEEEproof}

\begin{IEEEproof}[Proof of Lemma \ref{prison2}]
    It suffices to show that for any potentially attracting intersection point $p_{ij}$, $i\neq j$.
    Assume on the contrary that there exists a potentially attracting intersection point $p_{ii}$ that is intersected by $u^C_i$ and $u^D_i$ for some agent type $i$.
    From \eqref{eq3}, we know that when $n^C = n$, we have $u_i^C(n) = \textcolor{Blue}{R_i}$ and $u_i^D(n) = \textcolor{RawSienna}{T_i}$. When $n^C = 0$, we have $u_i^C(0) = \textcolor{Emerald}{S_i}$ and $u_i^D(0) = \textcolor{Red}{P_i}$. 
	Hence, $u_i^D(0)>u_i^C(0)$ and $u_i^D(n)>u_i^C(n)$ according to \eqref{PDConsdtraintOnPayoffs}.
	This implies that the $n^C(p_ii)<0$ or $n^C(p_ii)>n$,
	which contradicts the definition of potentially attracting intersection points. 
\end{IEEEproof}

\begin{IEEEproof}[Proof of Proposition \ref{p99}]
    We prove by contradiction.
    Assume on the contrary that the solution trajectory never reaches an equilibrium.
    Hence, in view of Lemma \ref{prop:2}, it reaches a fluctuation set $\O$ at some time and remains there afterwards. 
    Let $u^D_k, k\in\{1,\ldots,m\},$ denote the max defective line in Lemma \ref{prison}.
    Due to Lemma \ref{fluct_pt}, there are fluctuation points at the corresponding left and right walls of $\O$, and the solution trajectory will, hence, fluctuate at these walls.
    Because of the randomness of the activation sequence, a type-$k$ cooperator will become active during one of these fluctuations and switch to defection, making $u^D_k$ active.
    Being the max line at every $n^C$, the line will never become inactive afterwards, driving the solution trajectory to the extreme equilibrium at $n^C = 0$.
    This is a contradiction, completing the proof.
\end{IEEEproof}

\begin{IEEEproof}[Proof of Theorem \ref{th6}]
    The proof is similar to that of Proposition \ref{th_coord}.
\end{IEEEproof}

\begin{IEEEproof}[Proof of Lemma \ref{le10}]
    For any type-$i$ agent, we have $u^C_i(0) = \textcolor{Emerald}{S_i}n > u^D_i(0) = \textcolor{Red}{P_i}n$ and $u^C_i(n) = \textcolor{Blue}{R_i} < u^D_i(n) = \textcolor{RawSienna}{T_i}$. This leads to the proof.
\end{IEEEproof}

%%%
\bibliographystyle{IEEEtran}
  \bibliography{main}

% \begin{IEEEbiographynophoto}{Yiheng Fu}
% is currently working as a research assistant in the Centre for Mathematical Biology, Department of Mathematical and Statistical Sciences at the University of Alberta.

% \begin{IEEEbiographynophoto}{Pouria Ramazi}
% received the B.S. degree in electrical engineering in 2010 from University of Tehran, Iran, the M.S. degree in systems, control and robotics in 2012 from Royal Institute of Technology, Sweden, and the Ph.D. degree in systems and control in 2017 from the University of Groningen, the Netherlands. 
% He is currently a joint Postdoctoral Research Associate with the Departments of Mathematical and Statistical Sciences and Computing Sciences of the University of Alberta.

\end{document}